\documentclass[graybox]{style/svmult}

\usepackage{type1cm}        % activate if the above 3 fonts are
\usepackage{makeidx}         % allows index generation
\usepackage{graphicx}        % standard LaTeX graphics tool
\usepackage{multicol}        % used for the two-column index
\usepackage[bottom]{footmisc}% places footnotes at page bottom

\usepackage{newtxtext}       % 
\usepackage{newtxmath}       % selects Times Roman as basic font

\usepackage{ulem}
\usepackage[round,compress]{natbib}
\usepackage{amsmath}
\usepackage{BibDef}
\usepackage{url}

\makeatletter
\def\cleardoublepage{\clearpage}
\makeatother

\makeindex             % used for the subject index
\begin{document}

\title*{Multimessenger prospects of quasi-periodic eruptions}
\author{Vojtěch Witzany, Alessia Franchini, Matteo Bonetti, and Luca Broggi}
\institute{ Vojtěch Witzany \at Institute of Theoretical Physics, Faculty of Mathematics and Physics,
\and Alessia Franchini \at Dipartimento di Fisica “A. Pontremoli”, Universit{\`a} degli Studi di Milano, Via Giovanni Celoria 16, 20134 Milano, Italy. \email{alessia.franchini@unimi.it}
\and
 Matteo Bonetti \at Dipartimento di Fisica “G. Occhialini”, Universit{\`a} degli Studi di Milano-Bicocca, Piazza della Scienza 3, 20126 Milano, Italy. \email{matteo.bonetti@unimib.it}
 \and
Luca Broggi \at Dipartimento di Fisica “G. Occhialini”, Universit{\`a} degli Studi di Milano-Bicocca, Piazza della Scienza 3, 20126 Milano, Italy.% INFN, Sezione di Milano-Bicocca, Piazza della Scienza 3, I-20126 Milano, Italy. Removed from consistency - Luca
\email{luca.broggi@unimib.it}}
\maketitle

\abstract*{Quasi-Periodic Eruptions (QPEs) are recurring soft X-ray transients that may be generated by inspirals of stellar-mass objects spiraling into supermassive black holes, known as extreme mass ratio inspirals (EMRIs). Independently, EMRIs and the gravitational-wave signals they generate are one of the key targets for the Laser Interferometer Space Antenna (LISA). What is the potential of a coincident detection of EMRIs both as a QPE and by LISA? Electromagnetic counterparts to LISA events would provide sky localization, enable standard siren measurements of the Hubble constant and constrain formation mechanisms of the corresponding inspirals. Combined observations would link the accurate measurements of black hole masses and spins to their galactic nuclear environments and would thus enable lasting synergies with various observations across the electromagnetic spectrum. However, most of the currently known QPEs imply EMRI orbital periods that place the frequencies of the corresponding gravitational-wave signal out of the LISA sensitivity band. Additionally, the selection biases of QPE detections and the LISA instrument may preclude a coincident detection. Future searches should focus on expanding the QPE catalogs and ultimately hunt for ``golden'' short-period QPEs that correspond to EMRIs that fall within the LISA band.}

\abstract{Quasi-Periodic Eruptions (QPEs) are recurring soft X-ray transients that may be generated by inspirals of stellar-mass objects spiraling into supermassive black holes, known as extreme mass ratio inspirals (EMRIs). Independently, EMRIs and the gravitational-wave signals they generate are one of the key targets for the Laser Interferometer Space Antenna (LISA). What is the potential of a coincident detection of EMRIs both as a QPE and by LISA? Electromagnetic counterparts to LISA events would provide sky localization, enable standard siren measurements of the Hubble constant and constrain formation mechanisms of the corresponding inspirals. Combined observations would link the accurate measurements of black hole masses and spins to their galactic nuclear environments and would thus enable lasting synergies with various observations across the electromagnetic spectrum. However, most of the currently known QPEs imply EMRI orbital periods that place the frequencies of the corresponding gravitational-wave signal out of the LISA sensitivity band. Additionally, the selection biases of QPE detections and the LISA instrument may preclude a coincident detection. Future searches should focus on expanding the QPE catalogs and ultimately hunt for ``golden'' short-period QPEs that correspond to EMRIs that fall within the LISA band.}

%\tableofcontents

%%%%%%%%%%%%%%%%%%%%%%%%%%%%%%%%%%%%%%%%%%%%%%%%%%%%%%%%%%%%%%%%%%%%%
%%%%%%%%%%%%%%%%%%%%%%%%%%%%%%%%%%%%%%%%%%%%%%%%%%%%%%%%%%%%%%%%%%%%%
\section{Introduction: A new window into extreme mass-ratio systems}
%%%%%%%%%%%%%%%%%%%%%%%%%%%%%%%%%%%%%%%%%%%%%%%%%%%%%%%%%%%%%%%%%%%%%
%%%%%%%%%%%%%%%%%%%%%%%%%%%%%%%%%%%%%%%%%%%%%%%%%%%%%%%%%%%%%%%%%%%%%

\begin{figure}[htbp]
    \centering
    % --- LEFT COLUMN ---
    \begin{minipage}[c]{0.64\textwidth}
        \centering
        \includegraphics[clip, trim=1.35cm 0cm 1.35cm 0cm,width=\textwidth, keepaspectratio]{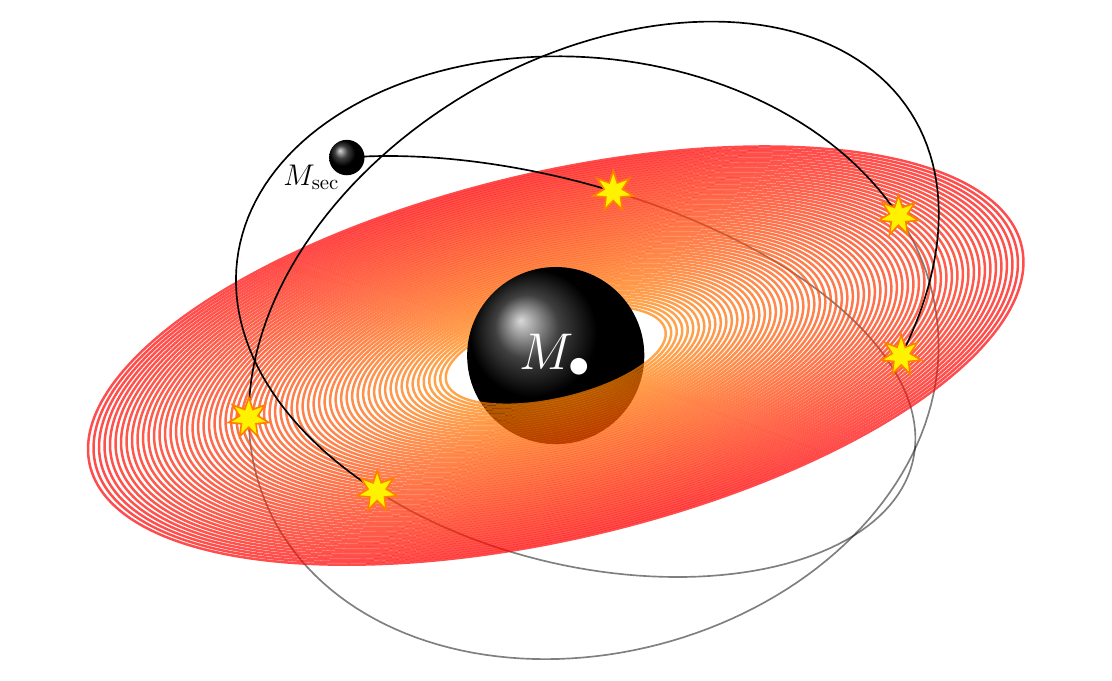}
        % This image will be vertically centered relative to the right column
    \end{minipage}
    \hfill
    % --- RIGHT COLUMN ---
    \begin{minipage}[c]{0.35\textwidth}
        \centering
        % Top PDF
        \includegraphics[width=\textwidth, keepaspectratio]{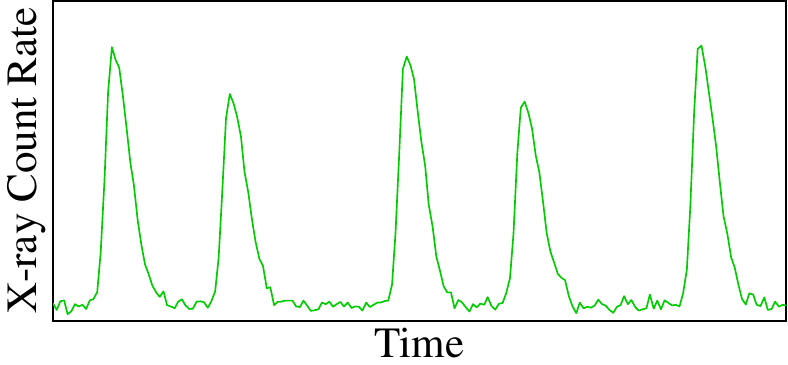}
        
        \vspace{0.cm} % Adjust gap between the two right-hand figures
        
        % Bottom PDF
        \includegraphics[clip, trim=0cm 0.1cm 0.1cm 0cm, width=\textwidth, keepaspectratio]{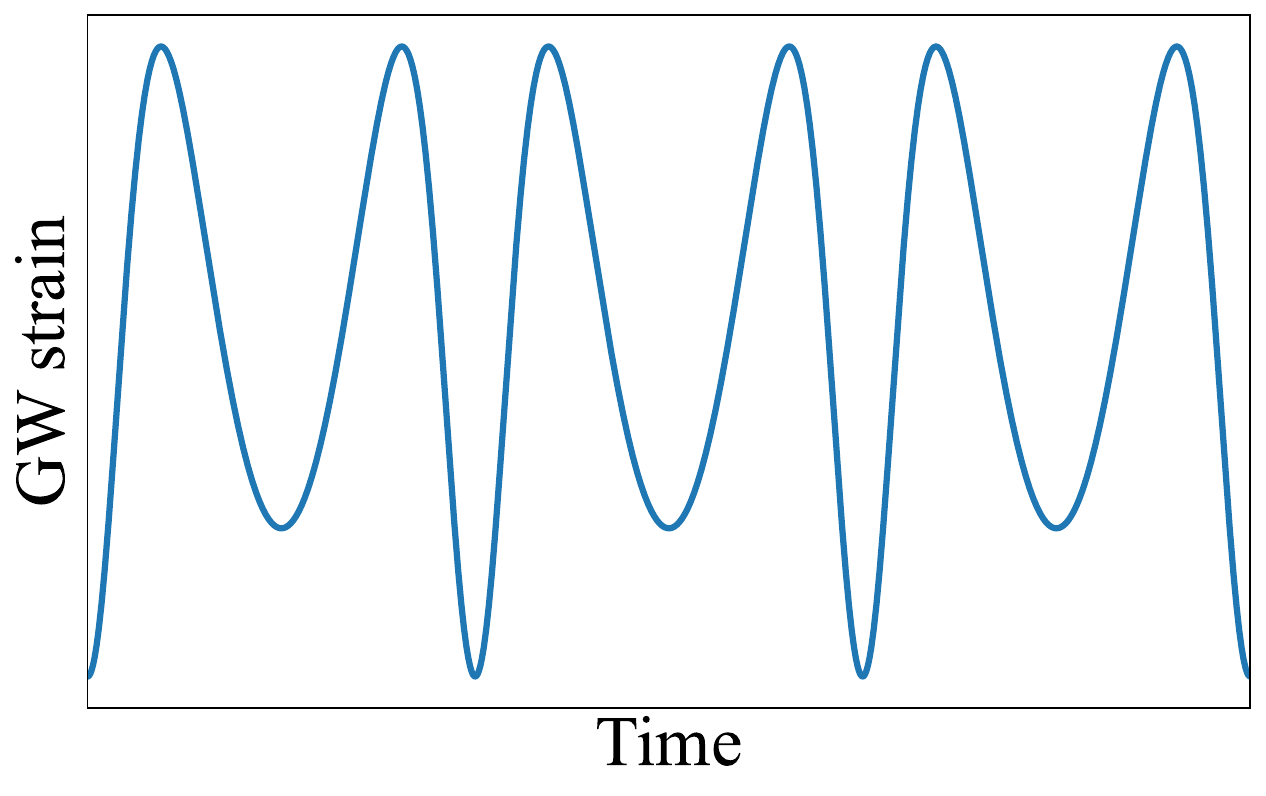}
    \end{minipage}

    \caption{\textit{Left:} A schematic view of the EMRI model of a QPE. A secondary object of mass $M_{\rm sec}$ orbits a MBH with mass $M_\bullet$ on a generally precessing, eccentric orbit. Its motion generates GWs, and collisions with the accretion disk lead to the soft X-ray eruptions. \textit{Right:} Mock signals from the two main messengers considered in this chapter: An X-ray lightcurve (green) and GW strain (blue) emerging from the process.}
    \label{fig:QPE-EMRI}
\end{figure}

As discussed in other chapters of this Handbook, observations by the X-ray telescopes XMM-Newton, Swift, Chandra, and eROSITA have revealed, among other transients, the existence of a new phenomenon known as Quasi-Periodic Eruptions (QPEs). 
First identified in the nuclei of nearby low-mass galaxies, QPEs manifest themselves as intense, recurring bursts of soft X-rays with periods ranging from hours to days \citep{Miniutti2019,Giustini2020,Arcodia2021,Chakraborty2021,Quintin2023,2024A&A...684A..64A,2024Natur.634..804N,HernandezGarcia2025}. 
Their host galaxies are often post-starburst systems, suggesting a link to recently faded nuclear activity \citep{2022A&A...659L...2W, 2024ApJ...970L..23W}. A possible connection to Tidal Disruption Events (TDEs) has been established through both statistical association and direct observation, with QPEs now confirmed to arise in the aftermath of optically-identified TDEs \citep{2024Natur.634..804N, 2025ApJ...983L..39C}. Furthermore, the recent observation of the ``Ansky" QPE in the turning-on AGN SDSS J1335+0728 suggests that QPEs may be more generally correlated with newly formed accretion disks \citep{HernandezGarcia2025}.

Several models have been proposed to explain QPEs. They range from instabilities in the accretion disk around the central massive black hole (MBH)  \citep{Raj2021,Pan2022,Kaur2023} 
to mass transfer models where the eruptions are powered by episodic accretion from a companion (either a star or a stellar-mass compact object) directly onto the MBH \citep{King2020, King2022, King2023, Metzger2022, Zhao2022, Chen2022, Krolik2022, Linial2023a, Yang2025, Lau2025}. 

The leading hypothesis, however, posits that QPEs are the electromagnetic signature of an Extreme Mass-Ratio Inspiral (EMRI) -- a stellar-mass object orbiting a massive black hole (MBH) -- that repeatedly interacts with a transient accretion disk, likely formed by the stellar debris of a tidally disrupted star \citep{2021ApJ...917...43S,2021ApJ...921L..32X,Linial2023b,2023A&A...675A.100F,Tagawa2023}. 
In this picture, the secondary object collides with the disk typically twice per orbit, generating shocks and outflows that give rise to the observed X-ray flares (see Fig. \ref{fig:QPE-EMRI}). The combination of different precession frequencies of both the disk and the secondary objects can qualitatively reproduce the different timing behaviors of QPEs.
Additional tests across a variety of observables, e.g., timing residuals \citep{2024A&A...690A..80A,Giustini2024,2024arXiv241100289P,2025A&A...693A.179M,2025ApJ...992..120C} and spectral energy density analysis \citep{Wevers2025,Guolo2025a,Chakraborty2025b}, are still ongoing to confirm the validity of these models.

If this EMRI connection is correct, QPEs could represent a milestone discovery as they would be the first electromagnetic counterparts to one of the key science targets for future space-based gravitational wave (GW) observatories, such as the adopted Laser interferometer space antenna (LISA) \citep{2017arXiv170200786A,2024arXiv240207571C}. In return, this would open a new frontier for multimessenger astronomy, promising to link the physics of accretion and electromagnetic radiation with the pure spacetime dynamics revealed by gravitational waves. This chapter provides an evaluation of these prospects, focusing on the capabilities of LISA and the scientific implications of detecting QPEs as electromagnetic counterparts to EMRIs.

This chapter is organized as follows. In Section~\ref{sec:GWIntro}, we review the fundamentals of gravitational wave astronomy and the LISA mission, and we describe the specific properties of Extreme Mass-Ratio Inspirals (EMRIs). We then examine the hypothesis that QPEs serve as electromagnetic counterparts to these events in Section~\ref{sec:QPE-EMRI-connection}, discussing current observational constraints and the population disparities that challenge the prospect of a combined detection of electromagnetic and GW signals. Section~\ref{sec:multi} outlines the transformative potential of multimessenger observations, detailing how joint detections could break parameter degeneracies, probe accretion disk physics and enable standard siren cosmology. Finally, we discuss the necessity of a coordinated search strategy to identify short-period ``golden'' QPEs and bridge the observational gap in Section~\ref{sec:future}.

%%%%%%%%%%%%%%%%%%%%%%%%%%%%%%%%%%%%%%%%%%%%%%%%%%%%%%%%%%%%%%%%%%%%%
%%%%%%%%%%%%%%%%%%%%%%%%%%%%%%%%%%%%%%%%%%%%%%%%%%%%%%%%%%%%%%%%%%%%%
\section{GW science and the LISA mission} \label{sec:GWIntro}
%%%%%%%%%%%%%%%%%%%%%%%%%%%%%%%%%%%%%%%%%%%%%%%%%%%%%%%%%%%%%%%%%%%%%
%%%%%%%%%%%%%%%%%%%%%%%%%%%%%%%%%%%%%%%%%%%%%%%%%%%%%%%%%%%%%%%%%%%%%

GW astronomy offers a fundamental complementary window to traditional electromagnetic observations, into the Universe. While electromagnetic radiation traces thermal and non-thermal emission from hot gas and accretion flows, gravitational waves directly encode the bulk motion of masses in strongly dynamical spacetimes. This makes GWs particularly valuable for studying compact objects in extreme environments where electromagnetic emission may be absent or intrinsically faint. Furthermore, GW detections are unimpeded by intervening matter, making them powerful probes across cosmic distances. In this Section, we intend to highlight the relevant aspects of the field of GW astronomy to readers more familiar with traditional astronomy in the electromagnetic bands; readers who are already well acquainted with LISA and EMRIs may wish to skip to Section \ref{sec:QPE-EMRI-connection}. We discuss elements of GW science in Section \ref{sec:GWbasics}, the LISA mission in Section \ref{sec:LISA}, and its source class that is central to our discussion -- EMRIs -- in Section \ref{sec:EMRIs}.

%%%%%%%%%%%%%%%%%%%%%%%%%%%%%%%%%%%%%%%%%%%%%%%%%%%%%%%%%%%%%%%%%%%%%
\subsection{Generation and detection of GWs} \label{sec:GWbasics}
%%%%%%%%%%%%%%%%%%%%%%%%%%%%%%%%%%%%%%%%%%%%%%%%%%%%%%%%%%%%%%%%%%%%%

A key observable for GW detection is the so-called space-time strain $h$ measured by observers far from a dynamical source. At leading order, and assuming that the distance to the observer is much larger than the size of the source, mildly relativistic sources produce the strain \citep[e.g.,][]{poisson2014gravity} 
\begin{align}
    h^{ij} = \frac{2 G}{c^4 D_{\rm L}} \ddot{I}^{ij}\,, \label{eq:htt}
\end{align}
where $\ddot{I}^{ij}$ is the second time derivative of the traceless mass quadrupole tensor of the source and $D_{\rm L}$ is the luminosity distance.  The wavelength of GWs is governed by the orbital period $P$ of matter orbiting in a source, $\lambda_{\rm GW} \sim 2 c/P$. In particular, the GW wavelength is almost always larger than the source region. 
GW emission is therefore a non-local effect that can be viewed as a ``toll'' paid by the system for conveying information about its changing matter configuration to the universe. 

An important observation to note about eq. \eqref{eq:htt} is that GWs are emitted by \textit{time-variable density contrasts}. As such, a stationary accretion disk near a stationary black hole may carry a non-zero mass quadrupole, but the generated GWs will be weaker by many orders of magnitude since the time variability of the mass quadrupole is small. Similarly, an orbiting hotspot or an eruption of an accretion disk may generate appreciable electromagnetic emission, but the effect on the mass quadrupole will typically be vanishingly small.  In particular, a characteristic mass $\mu$ moving with an orbital frequency $\Omega$ over characteristic orbital separations $R$ will correspond to a quadrupole $\ddot{I} \sim \mu \Omega^2 R^2 \sim \mu v_{\rm orb}^2$, where $v_{\rm orb}$ is the orbital speed. We can then estimate the characteristic strain as 
\begin{align}
    h \sim \frac{G \mu/c^2}{D_{\rm L}} \frac{v_{\rm orb}^2}{c^2}\,. \label{eq:hest}
\end{align}
We can identify in the above equation the quantity $G \mu/c^2$ as the gravitational radius of the mass $\mu$. 

For the sake of illustration, let us estimate an upper bound of the strain generated by a stationary accretion disk. If we use the standard accretion disk model of \citet{Shakura1973} with $\alpha = 0.1$, radiative efficiency $10\%$ and accreting close to the maximum Eddington accretion rate, then the mass of the \textit{entire} accretion disk with orbital periods up to some $P_{\rm max}$ can be estimated to have the mass
\begin{align}
    \mu_\mathrm{disk}(P<P_\mathrm{max}) \approx 2 \cdot 10^{-7} \left(\frac{M_\bullet}{10^6 M_\odot}\right)^{-1/3} \left(\frac{P_\mathrm{max}}{10 \, \mathrm{min}}\right) M_\odot\,,
\end{align}
where $M_\bullet$ is the central MBH mass. For the LISA detector, the natural cutoff is $P_\mathrm{max} \sim 10$ minutes (see Section \ref{sec:LISA}). According to eq. \eqref{eq:hest}, this entire mass orbiting in a single point around the nearest MBH in Sgr A$^*$ ($D_{\rm L} \sim 8$ kpc) at 10\% of the speed of light would generate a strain of $h \lesssim 10^{-26}$, orders of magnitude below the sensitivity of any current or planned detector. Similarly, the numerical study of \citep{Yuan:2025fde} demonstrated that even the stochastic GW signal generated by the ensemble of all accretion-disks around MBHs in the Universe is not strong enough to be detectable in the near future. From this, we see that GWs sufficiently strong for detection are predominantly generated by the motion of large amounts of mass. For distant, extragalactic sources of GWs, this leads to an inevitable selection bias towards heavy compact objects. 

Once emitted from the source, the interaction of GWs with matter is very weak and of a tidal nature. This is because the coupling of every matter particle to the strain $h^{ij}$ is universal, affecting matter collectively, at least for sizes below the GW wavelength. 
(This is unlike the case of electromagnetic waves, where the differences between charges lead to electromagnetic forces that pull apart particles at arbitrary separations down to the microscopic level.) 
As a result, current and planned GW detectors are predominantly \textit{passive}; their detection principle relies on the altered kinematics of the space-time through which the GWs are passing \citep[see, e.g.,][]{maggiore2008gravitational}. 

The key criterion for GW detection is the signal to noise ratio (SNR) 
obtained by projecting the time-series into a prediction of the signal, a method known as matched filtering \citep[e.g.,][]{1989CQGra...6.1761S,1998PhRvD..57.4535F}. While the immediate strain amplitude of most GW signals will not rise above detector noise during a single cycle, this method allows the SNR to accumulate over a number of cycles $N_{\rm c}$ as $\propto \sqrt{N_{\rm c}}$. For example, a continuous source at $\sim$1 mHz will accrue a total of $\sim 10^5$ cycles over a 4-year observation window, thus boosting its SNR by a factor of $\sim 10^2-10^3$ compared to its immediate signal strength. 

On the other hand, the matched filtering method places large computational requirements on the data analysis and requires  an \textit{a priori} understanding of the modeled systems. Additionally, the interpretation of the resulting SNR is subtle; the false alarm probability grows proportionally to the number of filtering trials, and low-SNR signals may be missed during the search in a large waveform space. This leads to an effective SNR cutoff, a threshold value for which the false alarm probability is sufficiently low and the signal is almost certain to be recovered by the search. The SNR cutoff is generally larger than 5 and depends on the characteristics of the detector, signal, and search algorithm.

Current and planned GW observatories span a wide range of detection frequencies, each sensitive to different astrophysical sources. Pulsar timing arrays (PTAs), such as NANOGrav, EPTA, and PPTA, leverage the precise timing of millisecond pulsars distributed across the galaxy to detect nanohertz gravitational waves ($f \sim 10^{-9}$--$10^{-7}$ Hz), primarily from supermassive black hole binaries and potentially from the stochastic GW background \citep[e.g.,][]{2010CQGra..27h4013H}. Ground-based laser interferometers operating in the audio-frequency band ($\sim$10 Hz--several kHz) include the LIGO-Virgo-KAGRA (LVK) network, which has detected hundreds of stellar-mass compact binary coalescences to date \citep{2025ApJ...995L..18A}. 

Looking towards the future, the planned Einstein Telescope (ET), a 10 km-scale underground triangular detector in Europe, aims to extend sensitivity down to $\sim$1 Hz while improving strain sensitivity by an order of magnitude \citep{punturo2010einstein}. Similarly, Cosmic Explorer (CE), a proposed 40 km-scale L-shaped detector in the United States, should operate in a comparable frequency range with further enhanced sensitivity \citep{2019BAAS...51g..35R}. However, nuclear transients repeating on the timescale from minutes to days, such as QPEs, occur at frequencies that are too low to be detectable by terrestrial detectors while also having frequencies that are too high to be detected by PTAs. The frequency gap between these detector classes will be filled at least in part by space-based detectors such as LISA, which we describe in the next section. 

%%%%%%%%%%%%%%%%%%%%%%%%%%%%%%%%%%%%%%%%%%%%%%%%%%%%%%%%%%%%%%%%%%%%%
\subsection{LISA: The millihertz GW observatory} \label{sec:LISA}
%%%%%%%%%%%%%%%%%%%%%%%%%%%%%%%%%%%%%%%%%%%%%%%%%%%%%%%%%%%%%%%%%%%%%

%%%%%%%%%%%%%%%%%%%%%%%%%%%%%%%%%%%%%%%%%%%%%%%%%%%%%%%%%%%%%%%%%%%%%
\begin{figure}
    \centering
    \includegraphics[width=1\linewidth]{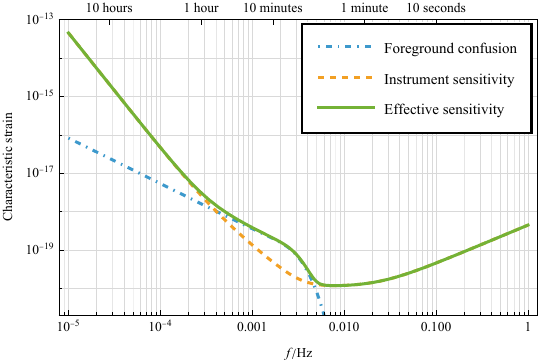}
    \caption{A logarithmic plot of expected LISA strain sensitivity as a function of frequency (bottom) or period (top), from \citet{2019CQGra..36j5011R} (see also \citet{2021arXiv210801167B}). Plotted is also the total effective sensitivity due to the foreground confusion caused by galactic compact binaries. LISA will be most sensitive to GWs sourced by orbital phenomena with $\sim$10-minute to 10-second periods. Sources outside of this range will have to be loud and/or persistent in order to be detectable.}
    \label{fig:sensitivity}
\end{figure}
%%%%%%%%%%%%%%%%%%%%%%%%%%%%%%%%%%%%%%%%%%%%%%%%%%%%%%%%%%%%%%%%%%%%%

LISA is a planned space mission led by the European Space Agency (ESA) designed to open a low-frequency GW window from approximately 0.1 mHz to 1 Hz \citep{2017arXiv170200786A,2024arXiv240207571C}. The mission was adopted by ESA and NASA in early 2024 and has moved to the industrial development phase as of June 2025. At the time of writing, it is on track to be launched into orbit in 2035.\footnote{As with any mission, timeline adherence depends on the stability of funding commitments from all international partners.} After approximately 2 years of cruise and commissioning after launch, it will start a 4.5 year science observation window in the late 2030s, with consumables that should allow for the extension of the observation window to up to a maximum of 10 years. 

The frequency range of LISA is inaccessible from the ground due to seismic noise. LISA will therefore target complementary classes of sources compared to terrestrial detectors. The frequency band is expected to be populated by a variety of sources ranging from compact white dwarf binaries within our galaxy to cosmological stochastic backgrounds \citep{AmaroSeoane2023}. However, the key targets for LISA are MBH binaries ranging from comparable-mass binaries to EMRIs with MBH masses within the $10^4-10^7 M_\odot$ range. 

In its very principle, LISA will operate as a laser interferometer similar to terrestrial GW detectors. It will consist of three spacecraft in a triangular formation with an arm length of 2.5 million kilometers, orbiting the Sun in an Earth-trailing orbit. 
Inside each spacecraft, a gold-platinum test mass will be in a state of near-perfect free-fall, shielded from non-gravitational forces such as solar radiation pressure and cosmic rays. 
The spacecraft will exchange laser beams, and the phase of the returning laser light will be compared to a local laser. A passing GW will stretch and squeeze the spacetime between the spacecraft, causing a minuscule change in the light travel time that will be  detected as a phase shift. However, the raw laser frequency noise in LISA will exceed the GW signal by many orders of magnitude but can be mitigated by time-delay interferometry (TDI), which linearly combines the six one-way inter-spacecraft phase measurements with appropriate time delays in post-processing \citep{2021LRR....24....1T}. This will effectively cancel the noise and reveal the underlying GW signal.

LISA's sensitivity will be governed by two main noise sources that cannot be eliminated by TDI \citep{2019CQGra..36j5011R,2021arXiv210801167B}. At low frequencies ($< 3$\,mHz), the sensitivity will be limited by the acceleration noise caused by residual non-gravitational forces acting on the test masses. At high frequencies ($> 3$\,mHz), the limiting factor will be shot noise in the photodetectors, which scales with the laser power. A further modulation will arise from LISA’s finite arm length, whose frequency-dependent arm response suppresses the sensitivity to GW signals at frequencies comparable to and above the light-travel frequency across the arms. Apart from instrumental effects, one should also expect that the population of galactic white dwarf binaries leads to source confusion, effectively lowering the sensitivity in some range of frequencies. The combination of these effects leads to the U-shaped sensitivity curve plotted in Figure \ref{fig:sensitivity}. The curve has an optimal ``sweet spot'' or ``bucket'' in the 1--10 mHz range -- a frequency band rich with astrophysical sources such as EMRIs.

%%%%%%%%%%%%%%%%%%%%%%%%%%%%%%%%%%%%%%%%%%%%%%%%%%%%%%%%%%%%%%%%%%%%%
\subsection{EMRIs among LISA primary sources} \label{sec:EMRIs}
%%%%%%%%%%%%%%%%%%%%%%%%%%%%%%%%%%%%%%%%%%%%%%%%%%%%%%%%%%%%%%%%%%%%%

As already mentioned in the Introduction, EMRIs are inspirals of stellar-mass objects into MBHs. EMRIs are slow, nearly adiabatic processes driven by the emission of GWs, lasting up to $10^4-10^5$ orbital cycles within the LISA band before the final plunge. 
This prolonged and intricate orbital evolution makes EMRIs exceptional probes of physics and astrophysics \citep{2017PhRvD..95j3012B}. Their detection through GW emission can provide crucial information on their formation channels \citep{2021PhRvD.103j3018P,2022MNRAS.514.3270B} and allow tests of general relativity \citep{2013LRR....16....7G}.

The GW signal from an EMRI is not a simple monotonic ``chirp'', that is, just a quasi-harmonic signal rising in frequency and amplitude. In fact, the EMRI waveform can be viewed as composed of many superposed harmonics chirping at different rates
\citep{2021PhRvD.103j4014H,WaveformWGLISA}. The orbit is a precessing trajectory that densely maps the spacetime geometry of the central MBH. The relativistic effects encoded in the waveform include periastron (apsidal) precession and Lense-Thirring (nodal) precession, which become non-perturbatively large or even divergent in the late stages of the inspiral. 

The characteristic frequency at which the EMRI GW signals peak
can be estimated from the orbital frequency of a particle at the innermost stable circular orbit of a Schwarzschild black hole with mass $M_\bullet$, which is given by
\begin{align}
    f_{\rm orb} = \frac{1}{2 \pi}\sqrt{\frac{G M_{\bullet}}{r_{\rm ISCO}^3}} = 2 \cdot 10^{-2}  \left(\frac{10^5 M_{\odot}}{M_\bullet}\right) \text{Hz} \,.
\end{align}
At this frequency %, the signal will approximately peak, 
the orbital motion will lose stability and the GW signal will terminate in a few cycles. As a result, LISA will be sensitive to EMRIs into MBHs with masses $M_{\bullet} \lesssim 10^7M_{\odot}$. The loudest EMRIs are expected to come from the ``sweet spot'' in the sensitivity curve, which corresponds to inspirals into MBHs with masses between $10^5-10^6 M_\odot$.

Since EMRIs complete many cycles in the LISA band, their signal phase must be modeled and recovered with extraordinary precision to allow for a very accurate measurement of the system's parameters \citep{WaveformWGLISA}. A single high-SNR EMRI detection is expected to yield extremely precise measurements for mass and spin: the \textbf{redshifted} masses of the central MBH and the orbiting compact object can be measured with a fractional precision of $\sim 10^{-4}$ while the MBH spin can be measured to a fractional precision of $\sim 10^{-3}$--$10^{-4}$ \citep{2017JPhCS.840a2021G}. 
This level of precision will enable unprecedented tests of fundamental physics. For instance, the no-hair theorem of General Relativity states that a Kerr black hole is uniquely described by its mass and spin and that all higher multipole moments are determined by these two parameters. An EMRI signal provides a direct map of these moments. By measuring the quadrupole moment independently and checking for consistency with the mass and spin, LISA can perform a percent-level test of the no-hair theorem \citep{2019PhRvD..99h4001D}.

Orbital parameters such as the eccentricity and inclination of EMRIs will be determined with similar precision, at least for EMRIs with negligible interactions with an accretion disk \citep{2024arXiv240207571C}, and with lower precision for disk-embedded EMRIs \citep{2024MNRAS.531.1506B}. While the mass of the light secondary will be measured to high precision, it is an open question whether sub-leading parameters such as the spin of the secondary can be constrained with EMRIs due to their small and possibly degenerate impact on the waveform \citep{2021PhRvD.104l4019P,2025PhRvD.111j3044C,skoupy2025}. Sky localization will be possible within a few square degrees, and the distance will be determined within a $\sim 10\%$ error. Consequently, one should not expect that EMRI host galaxies will be identified by the analysis of the LISA signal alone, even though statistical catalog matches could still provide useful cosmological information \citep{2021MNRAS.508.4512L,Lops_2023}. 

EMRIs also constitute a formidable data analysis challenge, as their low amplitude signals are buried deep within instrumental noise and the confusion-limited foreground from other astrophysical sources (primarily the stellar compact binaries in our Galaxy). A blind search for a general signal of $p$ parameters naively requires $\sim 10^p$ evaluations of waveform templates. The EMRI waveform space has at least 14 dimensions, making template-based methods extremely computationally intensive. As a result of such considerations, the SNR cutoff for the recovery of LISA EMRIs is empirically estimated to be between $15-20$ \citep{babak2010mock}. The value of the cutoff may be revised in the future since the development of efficient search and parameter estimation algorithms is an ongoing and active area of research \citep{2023PhRvD.107f3004L,2025PhRvD.111b4060K,2025PhRvD.111j3014D}.

%%%%%%%%%%%%%%%%%%%%%%%%%%%%%%%%%%%%%%%%%%%%%%%%%%%%%%%%%%%%%%%%%%%%%
%%%%%%%%%%%%%%%%%%%%%%%%%%%%%%%%%%%%%%%%%%%%%%%%%%%%%%%%%%%%%%%%%%%%%
\section{QPEs as GW progenitors and counterparts} \label{sec:QPE-EMRI-connection}
%%%%%%%%%%%%%%%%%%%%%%%%%%%%%%%%%%%%%%%%%%%%%%%%%%%%%%%%%%%%%%%%%%%%%
%%%%%%%%%%%%%%%%%%%%%%%%%%%%%%%%%%%%%%%%%%%%%%%%%%%%%%%%%%%%%%%%%%%%%

The EMRI model for QPEs might provide a concrete physical link between observable X-ray phenomenology and the emission of GWs. This section is dedicated to the specifics of this connection, exploring how the observed properties of QPEs can be used to test the model and what their GW signatures would be. In Section \ref{sec:testEMRIQPE} we discuss what constraints on the EMRI scenario can be obtained from existing observations. In Section \ref{sec:popdispar} we argue that coincident observations of 
QPE EMRIs by LISA might be unlikely and we further discuss the LISA detectability of a few known QPEs in Section \ref{sec:QPEGWdetection}. We finish with a brief discussion on the possibility of interpreting other quasi-periodic phenomena in galactic nuclei as EMRI signatures in Section \ref{sec:QPOunification}. 

%%%%%%%%%%%%%%%%%%%%%%%%%%%%%%%%%%%%%%%%%%%%%%%%%%%%%%%%%%%%%%%%%%%%%
\subsection{Testing the EMRI hypothesis with QPE observations} \label{sec:testEMRIQPE}
%%%%%%%%%%%%%%%%%%%%%%%%%%%%%%%%%%%%%%%%%%%%%%%%%%%%%%%%%%%%%%%%%%%%%

The validity of the EMRI model can be tested through continued X-ray monitoring of QPEs. 
If QPEs trace an orbit, their arrival times must follow the laws of relativistic dynamics. Long-term monitoring can reveal secular changes in the orbit due to GW emission and environmental interactions.
Furthermore, observed deviations from the predicted times of arrival of the flares can constrain the precession frequencies at play and, therefore, their underlying phenomena \citep{2025A&A...693A.179M}. 

%%%%%%%%%%%%%%%%%%%%%%%%%%%%%%%%%%%%%%%%%%%%%%%%%%%%%%%%%%%%%%%%%%%%%
\subsubsection{GW period decay} 
The most fundamental prediction is that the orbital period must decrease over time as the system loses energy to GWs. The leading estimate of period change $\dot{P}_{\rm GW}$ is a function of the system's masses and orbital separation and can be written as\footnote{We neglect the eccentricity enhancement of the period decay for the leading-order estimate since it does not change the order of magnitude for the moderate eccentricities inferred for QPEs.} \citep{Peters1964}
    \begin{equation}
        \dot{P}_{\rm GW} = -\frac{192\pi G^{5/2}}{5c^5}\frac{M_{\bullet}\,M_{\rm sec}M}{a^{5/2}}
    \end{equation}
    where $M_{\bullet},\,M_{\rm sec}$ and $M$ are the MBH, secondary and total mass respectively, $a$ is the semi-major axis of the EMRI. We can express this using the leading-order period $P = 2\pi/\sqrt{GM/a^3}$ and linear order in $M_{\rm sec}$ to obtain 
    \begin{equation}
        \dot{P}_{\rm GW} = -9\cdot 10^{-10} \left(\frac{M_\bullet}{10^6 M_\odot}\right)^{2/3}  \left(\frac{P}{10 \text{ hours}}\right)^{-5/3} \frac{M_{\rm sec}}{M_\odot} \,. \label{eq:PdotGW}
    \end{equation}
    The first discovered QPE source, which, consequently, also has the longest baseline of period observations, is GSN 069 observed since 2018 \citep{Miniutti2019,Miniutti2023a,2025A&A...693A.179M}. Conveniently, with a $\sim 9$ hour recurrence time it also sits approximately in the middle of the range of known QPEs and can currently be considered a ``typical'' QPE. For GSN 069, the expected period derivative from GW emission alone is tiny, $\dot{P}_{\rm GW} \lesssim 10^{-8}$ or less than a second over a baseline of a year, which is not observable over the current baseline of less than a decade. 
    
    What is the detection threshold for the period drift? Let us provide a simple order-of-magnitude estimate.  A single eruption in a typical QPE lasts around an hour, and we can conservatively estimate that eruptions need to shift by one full peak width or, in this case equivalently, by approximately ten percent of the period in order for the shift to be robustly detectable. This may seem overly conservative, but it is justified by the fact that many QPE observations exhibit evolving long-short patterns, gaps in data, or vanishing and reappearing flares that make the analysis of their secular evolution quite difficult. Assuming that we can observe a QPE for $10$ years, this leads to a $\dot{P}_{\rm threshold} \sim 10^{-5}$. For known QPEs, from equation \eqref{eq:PdotGW}, we see that unless the secondary is an intermediate mass black hole with $M_{\rm sec} \gtrsim 10^4 M_{\odot}$, we cannot detect the GW decay using the existing observational baselines we have for QPEs. 
    
    Can we hope to observe the decay over a longer baseline? Fundamentally, the QPE-EMRI scenario needs the accretion disk to continuously ``fuel'' the eruptions. However, TDE disks fade and cool with an approximate $\propto t^{-5/3}$ time dependence, possibly thinning out the fuel for QPEs and hindering our ability to survey their long-term period evolution. On the other hand, the phenomenological link between TDE fading and QPE luminosity evolution seems to be much more complicated than a simple proportionality (see e.g. \cite{Miniutti2023a}), making it difficult to provide robust predictions for long-term follow-up \citep{Arcodia2022,Miniutti2023a,2023A&A...674L...1M,Chakraborty2024,2024ApJ...963L..47P,2024arXiv241100289P}.  
    
%%%%%%%%%%%%%%%%%%%%%%%%%%%%%%%%%%%%%%%%%%%%%%%%%%%%%%%%%%%%%%%%%%%%%    
\subsubsection{Hydrodynamic drag} 

Interactions with an accretion disk can induce a much larger, potentially detectable period decay \citep{2024A&A...690A..80A}. 
    The hydrodynamic drag exerted on an object with radius $R$ punching through an accretion disk can be estimated as \citep{2021ApJ...917...43S,Linial2024a} 
    \begin{equation}
        \dot{P}_{\rm drag} \approx -\frac{3\pi R^2_{\rm sec}\,a}{M_{\rm sec}}\frac{\Sigma_{\rm d}(r)}{r} 
        \label{eq:hydrodrag}
    \end{equation}
    where $a$ is the semi-major axis of the EMRI, $R_{\rm sec}$ and $M_{\rm sec}$ are the radius and mass of the secondary
    , and $r$ is the radius within the disk at which the collision occurs. To account for compact objects, one can replace $R_{\rm sec}$ with the Bondi-Hoyle-Lyttleton radius of the compact object \citep[e.g.,][]{2021ApJ...917...43S}, which reads
    \begin{align}
        R_{\rm BHL} = \frac{2 G M_{\rm sec}}{v_{\rm rel}^2 + v_{\rm s}^2}\,, \label{eq:RBHL}
    \end{align}
    where $v_{\rm rel}$ is the relative speed between the disk and the secondary upon impact, and $v_{\rm s}$ is the sound speed in the disk. 
    For eRO-QPE2, assuming a star as the secondary object and a typical AGN disk, the period derivative was estimated to be $\sim 10^{-6}$ \citep{2024A&A...690A..80A}. This also falls below the detectability threshold, similarly to $\dot{P}_{\rm GW}$.

    For a compact object secondary, the hydrodynamic drag is essentially negligible unless its orbit is at a very low inclination with respect to the disk plane \citep[thus leading to a large $R_{\rm BHL}$, see][]{2023A&A...675A.100F}. Furthermore, the lack of AGN-characteristic emission lines in the spectra of QPE sources suggests that they do not currently host a standard AGN disk but are instead recently switched-off nuclei \citep{2022A&A...659L...2W,AgisGonzalez26}. Therefore, the secondary is more likely to be interacting with a disk formed from the tidal disruption of a star rather than an AGN disk. TDE disks have a mass budget bounded by the mass and angular momentum of the disrupted star, implying a surface density that decreases with orbital radius far more steeply than in AGN disk models \citep{Mummery2025}. Consequently, the lower surface densities imply that the hydrodynamic drag reduces steeply for longer-period QPEs compared to the estimates given above.

%%%%%%%%%%%%%%%%%%%%%%%%%%%%%%%%%%%%%%%%%%%%%%%%%%%%%%%%%%%%%%%%%%%%%
\subsubsection{Luminosity of QPEs as a function of masses and orbital parameters} \label{sec:luminosity}

The nature of the secondary object in the EMRI model of QPEs is unknown, and the mass of the primary MBH is typically constrained only within an order of magnitude from mass scaling relations \citep{2022A&A...659L...2W} or from TDE light-curve fitting \citep{Chakraborty2021,2024A&A...684A..64A,2025ApJ...983L..39C}. It is therefore difficult to eliminate the broad range of possible sub-scenarios within the EMRI scenario for QPEs. One of the few constraints on the nature of the secondary comes from the luminosity of the eruptions. 
According to most of the EMRI models for QPEs, the emission comes from two (one on each side of the disk) bubbles of optically thick gas that are pulled out of the disk as the secondary impacts it \citep{Linial2023b,2023A&A...675A.100F}. The exact initial size of these bubbles might differ depending on the assumed model, but it is somewhat related to the size of the colliding object \citep{1996ApJ...460..207L,1998ApJ...507..131I,2016MNRAS.457.1145P}. Since the eruptions are observed to be quasi-thermal, we provide an order-of-magnitude estimate for the emitted luminosity by assuming black-body radiation, which depends on the characteristic emission radius size $R_{\rm char}$ as
   \begin{align}
         L_{\rm QPE} \sim  3 \cdot 10^{42} \left(\frac{R_{\rm char}}{R_{\odot}}\right)^2 \left(\frac{T_{\rm QPE}}{10^6 K}\right)^4 {\rm erg \, s}^{-1}
     \end{align}
Given that the peak luminosities of QPEs are in the range of $10^{42}-10^{43} \rm{erg \, s^{-1}}$ and the eruptions temperatures are $\sim 10^6 K$, as they peak in the soft X-ray band, the characteristic radius must be $R_{\rm char} \sim  R_\odot$. 
    
More detailed models of the launching of the bubbles and the emission mechanism link $R_{\rm char}$ to the properties of the secondary \citep{Linial2023b,2023A&A...675A.100F}. Generally, the models set $R_{\rm char}$ as close to either the physical radius of the colliding body or the Bondi-Hoyle-Lyttleton radius (defined in eq. \eqref{eq:RBHL}), whichever is larger. This criterion essentially rules out lighter compact objects such as neutron stars or white dwarfs. Considering black holes as secondary objects, their effective Bondi-Hoyle-Lyttleton radius can reach the required size for either very massive objects $M_{\rm sec} \gtrsim 10^3 M_\odot$ or for lighter objects $M_{\rm sec} \sim 40-100 M_\odot$ if the orbit is sufficiently grazing with respect to the disc orbital plane. We further discuss $M_{\rm sec} \gtrsim 10^3 M_\odot$ secondaries in Section \ref{sec:imbh}.

%%%%%%%%%%%%%%%%%%%%%%%%%%%%%%%%%%%%%%%%%%%%%%%%%%%%%%%%%%%%%%%%%%%%%
\subsubsection{Precession induced modulations} 

The apsidal precession of the EMRI orbit itself can introduce super-orbital modulations into the QPE arrival times. Another source of super-orbital modulation is the Lense-Thirring induced nodal precession, of the EMRI orbit and potentially the disk, since the latter is likely to be formed by the tidal disruption of a star whose initial orbit is misaligned with respect to the MBH spin \citep{Stone2012}. 
While the disk nodal precession can be of the order of days \citep{Franchini2016} and therefore be probed by longer observational campaigns, the EMRI nodal precession period is much longer, therefore much harder to probe.
A tentative detection of a $\sim$19-day modulation in the timing residuals of GSN 069 has been interpreted as a signature of either rigid disk precession or the influence of a third body \citep{2025A&A...693A.179M, 2023A&A...675A.100F}. Confirming such modulations and matching them to physical models would provide strong evidence for the EMRI scenario and offer a viable path to constraining the MBH spin.

%%%%%%%%%%%%%%%%%%%%%%%%%%%%%%%%%%%%%%%%%%%%%%%%%%%%%%%%%%%%%%%%%%%%%
\subsubsection{Further constraints} 

Another important constraint that the timing properties of QPEs place is on the lifetime of these collisions. For instance, the source RX J1301 \citep{Giustini2020,Giustini2024} has been showing QPE flares since 2000, and the source GSN 069 has shown flares for at least 10 years \citep{Miniutti2019, Miniutti2023a,2023A&A...674L...1M}. 
In the scenario where the secondary object is a solar mass star, the repeated impacts will eventually lead to its complete disruption. The general scenario presented in the recent hydrodynamical simulation study of \citet{2025ApJ...978...91Y} is that every time the star flies through the disk, its outermost layer is stripped by ram pressure. Additionally, the impacts lead to shock heating of the remaining material of the star, and thus also to the expansion of its outermost layer. In return, the outer layer becomes less bound and mass loss per collision increases over time. In general, tidal forces further accelerate the stripping since the stars are often near their tidal radii (see Section \ref{sec:TDEcrit} and \citet{2023A&A...675A.100F}). This runaway process gradually ablates all of the material from the orbiting star.

How tightly does the stripping lifetime constrain known QPEs? For a solar mass orbiter and eRO-QPE2, \citet{2025ApJ...978...91Y} estimated the lifetime of the orbiting star to be within decades. For more general setups but still with solar-mass stars, the stars were estimated to fully disrupt within hundreds of years. These stripping times have been inferred for a solar-like star and might differ for other types of stars. Lower mass stars have lower binding energies, so we would expect them to have significantly lower stripping lifetimes. Their small cross-section would also make them hard to reconcile with QPE luminosities, as discussed in Section \ref{sec:luminosity}. The other end of the mass spectrum is somewhat more favorable, main-sequence giants have binding energies that can be up to an order of magnitude higher than solar-mass stars. As such, one can expect the ram pressure to release approximately an order of magnitude less material per impact. Given that their total mass is also an order of magnitude larger, one could expect that they survive the stripping for up to two orders of magnitude longer than solar-like stars. However, such estimates would have to be confirmed by simulations that take the nontrivial runaway mechanisms into account. Furthermore, stars with higher masses have a much shorter lifetime $\propto  M^{-3}_*$ and this time may be shorter than the time required to migrate towards the tight orbit consistent with QPE observations. Either way, the current observational baselines are not long enough to rule out solar-like stars due to the stripping timescale, but tensions may soon arise for short-period QPEs such as eRO-QPE2 as we accrue more data.

Finally, there seems to be a correlation between the energy of the QPEs and their period, which further challenges some of the existing models \citep{Mummery2025,Linial2025}.  The tentative fit to the energy-period data gives a dependence on the QPE period $P^{5/3}_{\rm QPE}$. However, one could also argue that there are two different classes of sources, separated by different orders of magnitude in orbital period.  Given the irregularities in subsequent peak profiles observed in the majority of QPE sources (e.g., RX J1301, eRO-QPE1), and the implied error bars on their energetics and mass of the central MBH, this correlation remains unclear.

%%%%%%%%%%%%%%%%%%%%%%%%%%%%%%%%%%%%%%%%%%%%%%%%%%%%%%%%%%%%%%%%%%%%%
\subsection{The disparity between LISA and QPE EMRI populations} \label{sec:popdispar}
%%%%%%%%%%%%%%%%%%%%%%%%%%%%%%%%%%%%%%%%%%%%%%%%%%%%%%%%%%%%%%%%%%%%%

While the possible connection between QPEs and LISA observations is frequently highlighted in the literature, a more detailed consideration complicates the case. As derived in Section \ref{sec:testEMRIQPE}, the power released in QPEs implies that the secondary must be able to eject gas within a radius $\gtrsim R_{\odot}$ from the accretion disk. 
This creates a fundamental tension with the population of detectable LISA EMRIs, which are expected to be predominantly inspirals involving compact objects such as neutron stars and stellar-mass black holes. This expectation arises because of i) the selection bias of LISA and ii) the fact that main sequence stars will mostly be tidally disrupted before reaching the LISA band. This could imply that there will be no overlap between LISA detections and QPEs. We further elaborate on these points in the sections below.  

%%%%%%%%%%%%%%%%%%%%%%%%%%%%%%%%%%%%%%%%%%%%%%%%%%%%%%%%%%%%%%%%%%%%%
\subsubsection{Heavy secondary bias for LISA EMRIs} \label{sec:heavybias}
%%%%%%%%%%%%%%%%%%%%%%%%%%%%%%%%%%%%%%%%%%%%%%%%%%%%%%%%%%%%%%%%%%%%%
In order to understand the selection bias for heavier secondaries in LISA EMRIs, one can refer to mass and distance scalings in the leading GW waveforms from \citet{Peters1964} and the SNR LISA estimates given by \citet{2019CQGra..36j5011R}. 
In the small mass ratio limit, the peak in the GW strain from a binary inspiral is reached around the last stable orbit and scales as $h_{\rm peak} \propto M_{\rm sec}/D_{\rm L}$. Additionally, the number of cycles the inspiral spends near the peak scales as $N_{\rm c} \propto M_{\bullet}/M_{\rm sec}$. The SNR can then be estimated as $\propto h_{\rm peak} \sqrt{N_{\rm c}} \propto \sqrt{M_{\rm sec} M_{\bullet}}/D_{\rm L}$. The sources will be detectable above an effective SNR$_{\rm cut-off}$. The detection volume $V_{\rm detect}$ will thus scale with the cube of the cut-off luminosity distance as
\begin{align}
    V_{\rm detect} \propto \frac{(M_{\rm sec} M_{\bullet})^{3/2}}{\text{SNR}_{\rm cut-off}^3}\,.
\end{align}
For example, an EMRI of a $\sim 50 M_\odot$ stellar-mass BH into a MBH of a given mass will be detectable in a volume that is 3 orders of magnitude larger than that of a $\sim 0.5 M_\odot$ white dwarf EMRI into a MBH of the same mass. Other selection effects play a role, such as the fact that heavier members of nuclear star clusters tend to sink towards the MBHs in galactic nuclei due to dynamical friction and other mass segregation effects \citep{hopman2006effect}. These selection effects lead to population analysis studies often completely neglecting the possibility of detecting EMRIs of objects lighter than a few solar masses.

%%%%%%%%%%%%%%%%%%%%%%%%%%%%%%%%%%%%%%%%%%%%%%%%%%%%%%%%%%%%%%%%%%%%%
\subsubsection{Tidal disruption criterion} \label{sec:TDEcrit}
%%%%%%%%%%%%%%%%%%%%%%%%%%%%%%%%%%%%%%%%%%%%%%%%%%%%%%%%%%%%%%%%%%%%%

A non-compact object of mass $M_{\rm sec}$ and radius $R_{\rm sec}$ can be estimated to be tidally disrupted when it reaches an orbital separation $r_{\rm t}$ from the MBH of the order
\begin{align}
    r_{\rm t}  \sim R_{\rm sec}\left(\frac{M_\bullet}{M_{\rm sec}}\right)^{1/3} \, . 
\end{align}
For an eccentric orbit, this will occur around the pericenter. We can then deduce that the orbital frequency of the last orbit before tidal disruption is
\begin{align}
    f_{\rm t} 
    = \frac{1}{2 \pi} \sqrt{\frac{GM_\bullet(1-e)^3}{r_{\rm t}^3}} 
    =  \frac{1}{2 \pi} \sqrt{\frac{G M_{\rm sec}(1-e)^3}{R_{\rm sec}^3}} 
    < \frac{1}{2 \pi} \sqrt{\frac{G M_{\rm sec}}{R_{\rm sec}^3}}
\end{align}
The last expression is independent of the orbital parameters and the mass of the primary, and it has the meaning of the dynamical frequency (or the inverse of the dynamical timescale) of the star. In other words, a star spiraling onto a MBH of any mass will be tidally disrupted before the orbital frequency reaches its dynamical frequency. By examining known mass and radius measurements of main-sequence stars, one sees that the dynamical timescale is typically of the order of $10^3$ seconds \citep[e.g.][]{southworth2014debcat}. Thus, a main-sequence star will be tidally disrupted near or below the lower edge of the LISA frequency band at $\sim 1$ mHz. In summary, to evolve across an appreciable stretch of the LISA band, the secondaries need to be compact objects such as white dwarfs, neutron stars, or black holes. However, we again stress the point from Section \ref{sec:testEMRIQPE} that white dwarfs and neutron star sizes might be too small to reproduce the observed QPE luminosities.

Given the above considerations, the possibility of detectable main-sequence star EMRIs is considered so narrow that it is generally neglected in the LISA science studies \citep[cf. ][]{AmaroSeoane2023}. 
On the other hand, known QPEs are so close (within a few hundred Mpc, or $z<0.1$) that some of them could, in principle, be detectable by LISA for sufficiently heavy secondaries (see section \ref{sec:QPEGWdetection}). In general, however, one should not conservatively expect that the LISA EMRI population will overlap with known QPEs.

%%%%%%%%%%%%%%%%%%%%%%%%%%%%%%%%%%%%%%%%%%%%%%%%%%%%%%%%%%%%%%%%%%%%%
\subsubsection{Intermediate mass black holes as secondaries?} \label{sec:imbh}
%%%%%%%%%%%%%%%%%%%%%%%%%%%%%%%%%%%%%%%%%%%%%%%%%%%%%%%%%%%%%%%%%%%%%

One special case that avoids the considerations above is the possibility that QPE secondaries are intermediate mass BHs (IMBHs) with masses $M_{\rm sec} \gtrsim 10^3 M_{\odot}$. As discussed in Section \ref{sec:testEMRIQPE}, such objects have sufficiently large Bondi-Hoyle-Lyttleton radii to reproduce QPE luminosities. They also survive deep in the primary MBH strong field, and their inspirals would produce high-SNR signals for LISA. However, even though the broad idea may be exciting for multimessenger astronomy, it is necessary to clearly assess the astrophysical likelihood of this scenario. 

The most direct constraint on an IMBH explanation of QPEs comes from the fact that no strong drift was observed in the periods of QPEs over the few-year baselines accrued to date; this limits the possible IMBH mass to $\lesssim 10^4 M_{\odot}$ (see Section \ref{sec:testEMRIQPE}). The range of IMBH masses between $10^3-10^4 M_\odot$ is quite narrow, but the IMBH scenario is not entirely excluded. Can we further constrain the astrophysical likelihood of this scenario? While IMBHs may form in dwarf galaxies or globular clusters and reach a MBH \citep[see, e.g.][for a review]{AmaroSeoane2023}, identifying all known QPEs with IMBHs poses several challenges. 

First, the formation rate of IMBHs is uncertain, and even more so the rate with which they merge with galactic MBHs because of the ``stalling'' of dynamical friction in the last tenth of a parsec before merger \citep{coleman2004intermediate}. In particular, the detailed modeling of both the formation and inspiral rates may reveal a very low probability of such events. Second, the GW inspiral time $t_{\rm GW}$ of such binaries is quite short. Using the leading \citet{Peters1964} formulas, we obtain
\begin{align}
    t_{\rm GW} = 5 \cdot 10^2 \text{ years} \left(\frac{P_{\rm orb}}{10 \text{ hours}}\right)^{8/3} \left(\frac{M_\bullet}{10^6 M_{\odot}}\right)^{-2/3} \left(\frac{M_{\rm sec}}{10^3 M_\odot}\right)^{-1} .
    \label{eq:tGW}
\end{align}
Given that the $\sim 15$ known QPEs are from a nearby volume of the size of a few hundred Mpc (redshift $z \lesssim 0.05$), this would imply very high formation rates of MBH-IMBH binaries and, among other things, a quick mass growth of the central MBHs \citep[see also][]{Linial2023b}. 

We can turn these heuristics into a quantitative likelihood estimate. Let $R_{\rm m}$ be the mean IMBH inspiral rate per MBH host. The mass-growth budget of the host caps this at
\begin{align}
    R_{\rm m} \;\lesssim\; R_{\rm m}^{\rm max} \;\sim\; \frac{M_\bullet}{M_{\rm sec}\,t_{\rm H}} \;\sim\; 7\times 10^{-8}\;{\rm yr^{-1}\,host^{-1}}\,,
\end{align}
for $M_\bullet=10^{6}\,M_\odot, M_{\rm sec} = 10^3 M_\odot$ and a Hubble time $t_{\rm H}\simeq 1.4\times 10^{10}$\,yr. Let $n_{\rm MBH}\simeq 10^{-2}\,{\rm Mpc^{-3}}$ be the comoving number density of host MBHs in the relevant mass band \citep{2012MNRAS.421..621B}, $V_{\rm survey}\simeq (4\pi/3)(250\,{\rm Mpc})^{3}$ the volume containing the known short-period QPE sample, $f_{\rm disk}$ the fraction of MBHs whose discs are dense enough to fuel QPEs, and $f_{\rm compl}^{\rm short}$ the survey completeness for sources with $P_{\rm QPE}\le 10$\,h. The expected number of short-period detections is then
\begin{align}
    \Lambda \;\sim\; n_{\rm MBH}\,V_{\rm survey}\,f_{\rm disk}\,f_{\rm compl}^{\rm short}\,R_{\rm m}\,t_{\rm GW}(P_{\rm orb,max})
    \;\sim\; 0.60\,\left(\frac{f_{\rm disk}}{10^{-2}}\right) \left(\frac{f_{\rm compl}^{\rm short}}{0.45}\right)\,\,,
    \label{eq:LambdaIMBH}
\end{align}
where the upper bound is reached at $R_{\rm m}=R_{\rm m}^{\rm max}$, with $P_{\rm orb,max}=2\times 10\,{\rm h}$ identified from the impact-model assumption of two flares per orbit and $t_{\rm GW}(P_{\rm orb,max})\simeq 2.9\times 10^{3}$\,yr from Equation~\eqref{eq:tGW}. We focus on the short-period bin because long-period ($\sim 10$\,d) QPEs require weeks of cadenced monitoring that few campaigns have achieved making it very hard to estimate the corresponding $f_{\rm compl}^{\rm long}$. Among these sources, one could postulate a subpopulation of QPEs with IMBH as companions due to their longer GW inspiral time but the heavy systematics on $f_{\rm compl}^{\rm long}$ make the exploration of such hypotheses too challenging.

The short-period completeness $f_{\rm compl}^{\rm short}$ is itself difficult to derive from first principles: QPEs have been mostly discovered serendipitously through follow-up of known transients or archival reanalysis by pointing instruments such as \textit{Chandra} or \textit{XMM-Newton}. For such discoveries, one should assume the survey completeness to be a small fraction of unity. Only the scans of the X-ray sky by \textit{eROSITA} can be considered as a quasi-systematic search \citep{eROSITA-catalogue,Arcodia2021,2024A&A...684A..64A}.

Specifically, eROSITA completed 4 all-sky surveys, each separated by approximately six months. During each survey, a typical sky position was scanned approximately 6 times with consecutive scans separated by $\sim 4$-hour intervals, accumulating $\sim 40$ seconds of exposure per scan. While this provides a $\sim 90 \%$ chance to pick up on the variability of short-period sources with duty cycles around $10\%$, the identification of a QPE was only possible if it was picked for follow up by other instruments. The likely QPE candidates were picked by criteria such as \textit{repeating} and/or significant variability over the course of the surveys, X-ray softness, and correlations with other observations as self-analyzed by \citet{arcodia2024cosmic}. Another important point is that only data from the Western hemisphere were reported on so far \citep{eROSITA-catalogue}, putting a hard limit on survey completeness to $<0.5$. \citet{arcodia2024cosmic} then estimated the overall survey completeness to be within $0.1-0.3$ for the sources of the type discovered by eROSITA.

The short-period bin includes eRO-QPE2, but also previously discovered sources RX\,J1301.9+2747 and GSN 069. Do these add to survey completeness when added to eROSITA discoveries? RX\,J1301.9+2747 is in the Eastern hemisphere and thus increases completeness independently of the eROSITA sources. On the other hand, the source GSN 069 is in the Western hemisphere and would very likely be picked for follow-up; as such it does not increase eROSITA completeness. Assuming the two sources in the Western hemisphere represent a $\lesssim 0.3$ complete set of the whole (or $\lesssim 0.6$ just on the Western hemisphere), and that serendipitous discoveries across Eastern and Western hemispheres appear at half the rate in this bin, we estimate the total completeness $f_{\rm compl}^{\rm short} \lesssim 0.45$. However, we keep the dependence on $f_{\rm compl}^{\rm short}$ in equation~\eqref{eq:LambdaIMBH} explicit, so the result can be computed for any plausible value. We treat the three short-period sources (eRO-QPE2, RX\,J1301.9+2747, GSN\,069) as a single population $\mathcal{H}_{\rm IMBH}$ of IMBH-MBH inspirals: their luminosities, eruption profiles, and host-MBH masses are too similar to defend a multimodal population hypothesis.
 
The disk-bearing fraction $f_{\rm disk}$ is the dominant uncertainty. Using the central TDE rate of $10^{-5}-10^{-4}\,{\rm yr^{-1}\,galaxy^{-1}}$ inferred from observed properties \citep{StoneMetzger2016,Chang2024} and a fiducial $\sim 10$\,yr TDE-disc lifetime gives $f_{\rm disk}\sim 10^{-4}-10^{-3}$ for TDE-fuelled discs alone, with low-luminosity AGN contributing comparably. A larger overlap between these populations is not excluded -- some changing-look AGN may themselves be TDEs -- and forthcoming LSST samples will sharpen this estimate. We additionally note the argument of \citet{Mummery2025} that TDE-disc densities are too low to produce observable IMBH-driven flares in the first place 
, which would push the effective $f_{\rm disk}$ for the IMBH scenario further down. 
 
\begin{figure}
    \centering
    \includegraphics[width=0.85\linewidth]{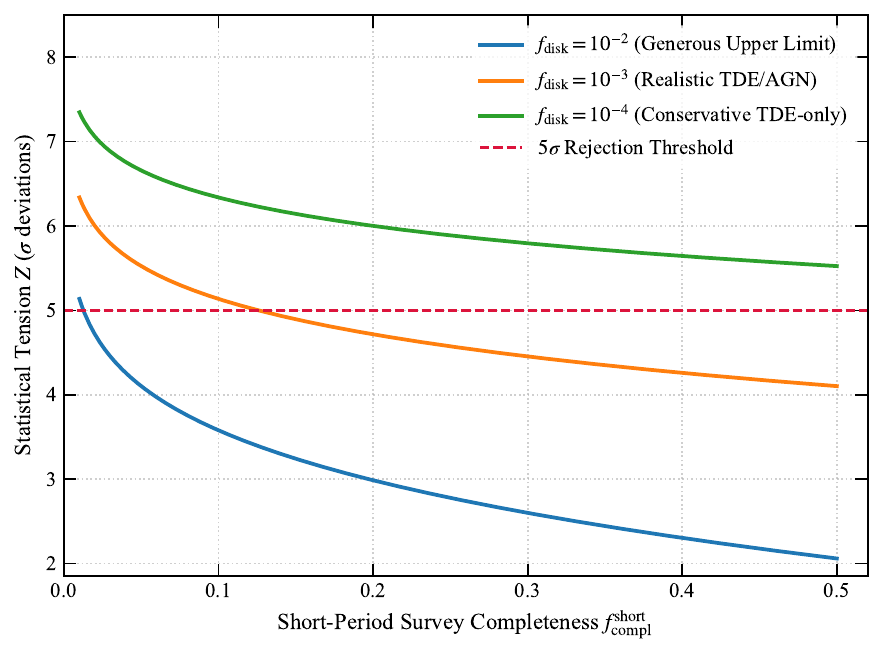}
    \caption{$\sigma$-tension of the hypothesis that all three short-period QPEs are IMBH-MBH inspirals, as a function of the short-period survey completeness $f_{\rm compl}^{\rm short}$, for three values of the disk-bearing fraction $f_{\rm disk}$. The dash-dotted line marks the conventional $5\sigma$ rejection threshold.}
    \label{fig:imbhlikelihood}
\end{figure}

Given $N_{\rm obs}=3$ short-period detections, the Poisson likelihood of $\mathcal{H}_{\rm IMBH}$ is $\mathcal{L}=\Lambda^{N_{\rm obs}} e^{-\Lambda}/N_{\rm obs}!$, with its maximum over $\Lambda$ at $\Lambda=N_{\rm obs}$, so that $\mathcal{L}_{\rm max}=27\,e^{-3}/3!\simeq 0.224$. We can then express the tension in terms of significance $Z$ (number of standard deviations $\sigma$ away from the mean) computed by Wilks' theorem
\begin{align}
    Z(\Lambda) = \sqrt{\,-2\ln(\mathcal{L}/\mathcal{L}_{\rm max})\,} = \sqrt{\,2[\,\Lambda - N_{\rm obs} + N_{\rm obs}\ln(N_{\rm obs}/\Lambda)\,]\,}\,.
\end{align}
$Z$ is shown as a function of $f_{\rm compl}^{\rm short}$ in Fig.~\ref{fig:imbhlikelihood}. Even at the highly generous corner that stacks assumptions favorable for the IMBH explanation of QPEs $(f_{\rm disk}=10^{-2},\,f_{\rm compl}^{\rm short}=0.45,\,R_{\rm m}=R_{\rm m}^{\rm max})$, the hypothesis already sits at $\Lambda\sim 0.60$, a $\simeq 2\sigma$ tension with $N_{\rm obs}=3$. Less favourable disk-bearing fractions $f_{\rm disk}\sim 10^{-3}$, and completeness $f_{\rm compl}^{\rm short}\lesssim 0.15$, drive the tension to $\sim 5\sigma$; eccentric orbits, factoring in disk-fed MBH growth, dynamical-friction stalling, or one impact per orbit would each tighten the bound further. We conclude that by this simple estimate (including a number of steps valid only at the level of orders of magnitude) the IMBH-only interpretation of the short-period QPE population is in significant tension across the plausible parameter range. A more precise statistical argument should be carried out in a dedicated study and is out of scope for this review.
 
%%%%%%%%%%%%%%%%%%%%%%%%%%%%%%%%%%%%%%%%%%%%%%%%%%%%%%%%%%%%%%%%%%%%%
\subsection{LISA detection prospects for known QPEs} \label{sec:QPEGWdetection}
%%%%%%%%%%%%%%%%%%%%%%%%%%%%%%%%%%%%%%%%%%%%%%%%%%%%%%%%%%%%%%%%%%%%%

It is instructive to quantify the LISA detection horizon for specific EMRI scenarios corresponding to known QPE sources, even though some of the scenarios may be unlikely given the astrophysical and observational arguments discussed in Section \ref{sec:popdispar}. Among the key parameters that establish detectability are the primary MBH mass, the orbital frequency (inferred from the observed QPE period), and the distance of the source. Additionally, detectability is a strong function of the orbiter's mass, scaling roughly with $\sqrt{M_\mathrm{sec}}$ (see Section \ref{sec:heavybias}). 

A recent study by \citet{Zhan2026} conducted an analysis of the known QPE population. They assumed the secondaries to be either white dwarfs, stellar mass BHs with mass $100 M_\odot$, or intermediate-mass BHs with masses $10^3 M_{\odot}$. In the case of white dwarfs and stellar-mass BHs, none of the known QPEs were found to be detectable (SNR below 10); for intermediate mass BHs, both eRO-QPE2 and eRO-QPE4 could be robustly detectable (SNR$>20$) by LISA with a 4-year observation campaign. The IMBH scenario represents the only possibility for the currently known QPEs to generally fall within the LISA sensitivity band. However, since this scenario is also strongly astrophysically disfavored (see Section \ref{sec:imbh}), we conclude that currently known QPEs are generally not expected to have GW counterparts detectable by LISA.

 Nevertheless, there is a caveat to the study of \citet{Zhan2026}. In this work, the authors considered only zero-eccentricity orbits in their modeling of the GW signal. However, eccentricity produces more GW power in higher frequency harmonics, which are closer to the LISA sensitivity band and may enhance the detectability of the signal. For example, RX J1301.9+2747, with a period of $\sim$ 4 hours, could be detectable if the secondary object is a stellar-mass BH of $>35\,M_\odot$ and its orbit has a moderate eccentricity of $e \approx 0.25$ \citep{2025arXiv250807961L}.\footnote{We reiterate that a stellar-mass BH would have to be on a low-inclination orbit to reproduce QPE luminosities, see Section \ref{sec:luminosity}.}
 
\begin{figure}
    \centering
    \includegraphics[width=0.8\linewidth]{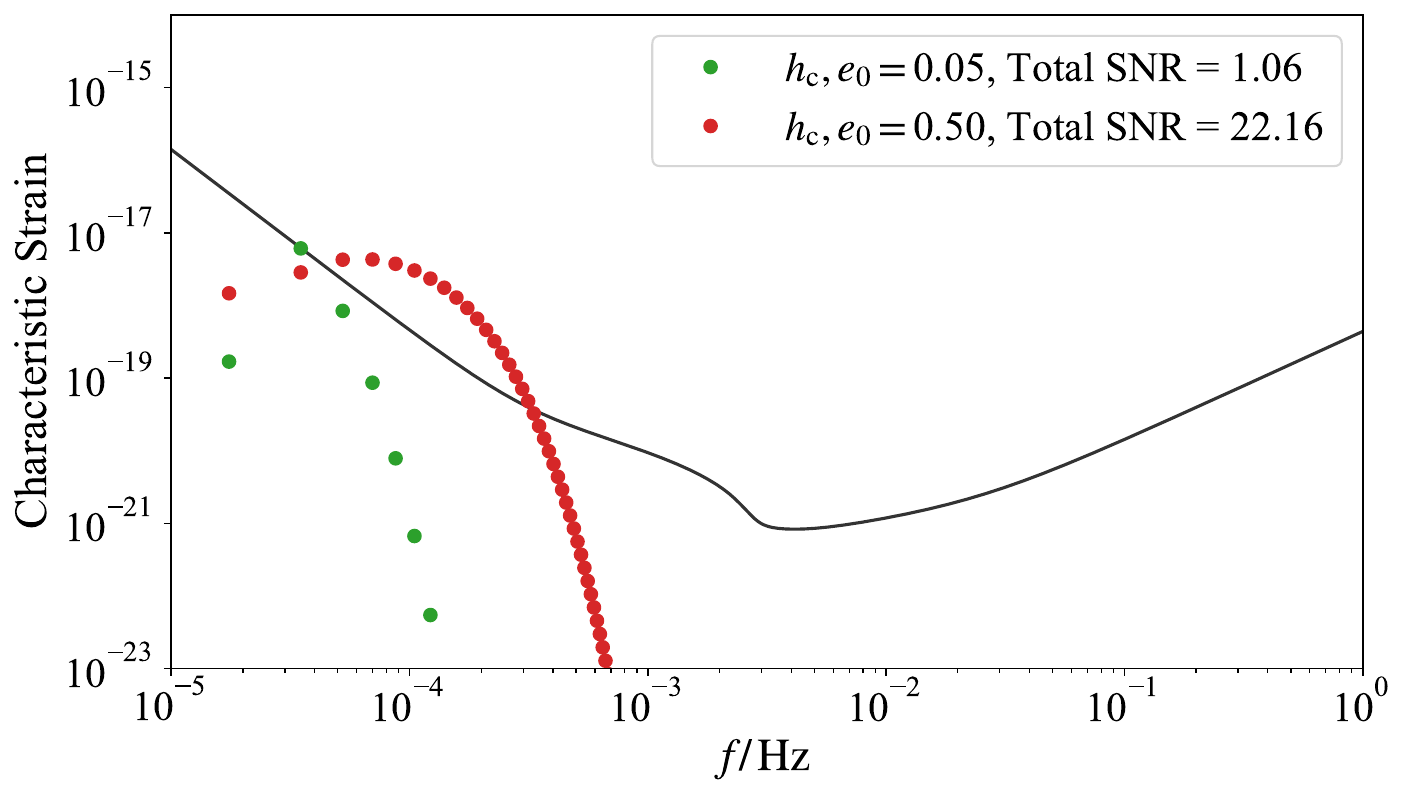}
    \caption{Characteristic strain (green and red) for an EMRI with $M_\bullet = 4.47 \cdot 10^6 \  M_\odot$, $M_{\rm sec} = 30 \ M_\odot$, $D_L = 100 \ \rm Mpc$ and $a = 55.5 R_g$ and two different initial eccentricities $e_0$ as indicated on the plots. The points correspond to the contribution of individual frequency harmonics of the waveform and a point above the LISA sensitivity curve (black) signifies a contribution to SNR that is larger than 1 \citep[see][for more]{2019CQGra..36j5011R}. Assuming a higher eccentricity leads to more power in the higher harmonics, which are closer to LISA sensitivity band. The physical parameters correspond to possible parameters of the EMRI generating the QPE in RX J1301.9+2747 and were selected following \citet{2025arXiv250807961L}.}
    \label{fig:RXJstrain}
\end{figure}

 We illustrate this point by plotting the characteristic strain produced by orbits with initial eccentricities $e_0 =0.05$ and $e_0 = 0.5$ using the leading-order waveform model from \citet{peters1963gravitational} and \citet{Peters1964} in Fig. \ref{fig:RXJstrain}. Because the EMRI corresponding to RX J1301.9+2747 would be in the very early inspiral stage, its frequency would essentially not change during the LISA mission lifetime, and the plotted GW strain is effectively restricted to points in the frequency spectrum. However, as mentioned, eccentric orbits also produce GW power at higher harmonic frequencies, which leads to the appearance of separate points corresponding to the strain contributed by each harmonic. From this, we see the reason why eccentricity is important: even though the same underlying orbital period and thus fundamental GW frequency is assumed in both cases, the higher-eccentricity signal peaks at four times the fundamental frequency, while the low-eccentricity case peaks at the second harmonic. Additionally, the higher eccentricity produces a broader distribution of GW power that overlaps significantly with the LISA sensitivity ``bucket''.

We would like to stress that the blind search for EMRIs may not be optimized for finding signals such as the GW counterpart of RX J1301.9+2747. This is because the slowly chirping harmonics of the early-inspiral signal could easily be confused one by one with individual galactic binaries. As such, even a GW signal with a cumulative SNR of 20 may be difficult to disentangle and interpret correctly in the full LISA datastream due to the concurrent overlap of many GW signals.  

Finally, we also emphasize the importance of an eventual non-detection of a GW signal from a well-characterized QPE source. A null detection still provides valuable astrophysical information, most importantly constraints on the mass of the orbiting object. For example, an upper bound on the $M_{\rm sec}$ below $\sim 0.1 M_{\odot}$
for a luminous QPE would make it difficult to reproduce the observed luminosity in any version of the QPE-EMRI scenario. In return, this would favor different explanations that do not produce any detectable GW signals, such as disk instabilities.

%%%%%%%%%%%%%%%%%%%%%%%%%%%%%%%%%%%%%%%%%%%%%%%%%%%%%%%%%%%%%%%%%%%%%
\subsection{GW detectability in a unified model of quasi-periodic phenomena?} \label{sec:QPOunification}
%%%%%%%%%%%%%%%%%%%%%%%%%%%%%%%%%%%%%%%%%%%%%%%%%%%%%%%%%%%%%%%%%%%%%

QPEs fall within the broader class of repeating nuclear transients, which also includes quasi-periodic oscillations (QPOs), repeating TDEs, or quasi-periodic outflows \citep[see][for a recent review]{2025ARA&A..63..379K}. While QPOs can be explained by models invoking various accretion disk oscillations or orbiting hotspots \citep[e.g.][]{2013LRR....16....1A,2014SSRv..183...43B}, the recent emergence of QPEs prompts the question of whether low-frequency QPOs in galactic nuclei could also be caused by secondary objects perturbing the disk \citep{Kejriwal2024}. The general idea is that while QPEs might be caused by abrupt ``punching'' through the accretion disk, QPOs would correspond to the smooth modulations of the light-curve induced or emitted by an object embedded in the disk. Distinguishing between these regimes depends on whether the secondary's orbit allows for supersonic shocks or continuous subsonic interactions with the accretion flow. If this unified model of quasi-periodic phenomena in galactic nuclei holds, the existing catalog of X-ray QPOs would provide an additional set of pre-identified targets for LISA, offering a pathway to verify their nature through the detection or non-detection of coincident GWs.

A recent study by \citet{Kejriwal2024} explored this unification by modeling both QPEs and QPOs as potential electromagnetic counterparts to LISA-detectable EMRIs. By analyzing a population of EMRIs retrofitted with observational cutoffs (luminosity distance $D_L \le 1$ Gpc and LISA SNR $\ge 15$), they estimated that the mean frequency of quasi-periodic phenomena observable today should lie around $0.46 \pm 0.22$ mHz. This frequency band notably encompasses several well-known X-ray QPOs, suggesting that they are an interesting targets for further investigation.

As a demonstration, \citet{Kejriwal2024} modeled the well-known QPO source RE J1034+396 as an EMRI. This active galactic nucleus exhibits a $\sim 1$ hour periodicity in the soft X-ray band with evidence for a significant decrease in the QPO period. 
Assuming that the central MBH has a mass of $10^6 - 10^7 M_{\odot}$ and that the period shortening is solely due to GW emission of the secondary driving the QPO, their prior-predictive analysis estimated the secondary mass between $6 - 56 M_{\odot}$.
Crucially, such a system would be a possible target for LISA, with a predicted SNR of $\approx 14$, highlighting the potential for these high-frequency QPOs to serve as multimessenger EMRI candidates.

The observational landscape of millihertz QPOs extends beyond RE J1034+396. \citet{2019Sci...363..531P} reported a stable 131-second ($7.65$ mHz) X-ray QPO in the tidal disruption event (TDE) ASASSN-14li. The high frequency and stability of this signal imply an origin in the innermost regions of the accretion flow, very close to the event horizon of a rapidly spinning black hole (dimensionless spin $\gtrsim 0.7$). While the mechanism remains debated, the association of such stable periodicities with TDEs reinforces the link between debris disks and repeating soft X-ray transients.

However, the interpretation of QPOs as simple EMRI chirps faces challenges when confronted with the frequency evolution observed in other sources. \citet{Masterson2025} discovered a high-significance millihertz QPO in the AGN 1ES 1927+654, which appeared following a major outburst and coronal destruction event. The QPO period decreased from $\sim 18$ minutes to $7.1$ minutes over two years, a rapid evolution that is difficult to reconcile with a pure GW inspiral, as the frequency derivative ($\dot{f}$) was observed to decelerate ($\ddot{f} < 0$). Such behavior disfavors a standard GW chirp but may be consistent with models involving stable mass transfer from a Roche-lobe-filling white dwarf companion or complex disk-tearing instabilities.

%%%%%%%%%%%%%%%%%%%%%%%%%%%%%%%%%%%%%%%%%%%%%%%%%%%%%%%%%%%%%%%%%%%%%
%%%%%%%%%%%%%%%%%%%%%%%%%%%%%%%%%%%%%%%%%%%%%%%%%%%%%%%%%%%%%%%%%%%%%
\section{Multimessenger science with QPE-EMRIs} \label{sec:multi}
%%%%%%%%%%%%%%%%%%%%%%%%%%%%%%%%%%%%%%%%%%%%%%%%%%%%%%%%%%%%%%%%%%%%%
%%%%%%%%%%%%%%%%%%%%%%%%%%%%%%%%%%%%%%%%%%%%%%%%%%%%%%%%%%%%%%%%%%%%%

As discussed in the previous section, a GW detection of a QPE counterpart might be difficult or even impossible, depending on the details of the orbital configuration and the nature of the secondary. Still, if such a detection is achieved, it would allow for a synergy of the joint electromagnetic and GW observations, enabling science that is impossible with either messenger alone. Below, we list some achievements that would become feasible under the assumption of coincident electromagnetic and GW detections of QPE-EMRIs. In Section \ref{sec:EMRIenviron}, we discuss the disentangling of the various parameters of the source brought about by a multimessenger detection. In particular, these would allow us to constrain accretion disk properties, which we comment on in Section \ref{sec:accretiondisk}; EMRI formation channels, which are discussed in Section \ref{sec:EMRIformation}; and to provide an independent measurement of the Hubble constant, which we discuss in Section \ref{sec:sirens}.

%%%%%%%%%%%%%%%%%%%%%%%%%%%%%%%%%%%%%%%%%%%%%%%%%%%%%%%%%%%%%%%%%%%%%
\subsection{EMRI parameters} \label{sec:EMRIenviron}
%%%%%%%%%%%%%%%%%%%%%%%%%%%%%%%%%%%%%%%%%%%%%%%%%%%%%%%%%%%%%%%%%%%%%

The most transformative aspect of a multimessenger QPE-EMRI detection lies in breaking the fundamental degeneracies inherent to each messenger in isolation. While LISA detects the redshifted masses $\mathcal{M}_z = (1+z)\mathcal{M}$ with high precision, the vacuum gravitational-wave signal alone cannot decouple the source redshift $z$ from the intrinsic black hole mass $M_{\bullet}$. A spectroscopic redshift from the QPE host galaxy breaks this degeneracy. Even for sources at the LISA detectability threshold (SNR $\approx 20$), the high number of orbital cycles ($\sim 10^5$) allows the source-frame MBH mass to be determined with a relative precision of $\Delta M_{\bullet}/M_{\bullet} \sim 10^{-3}$ \citep{2017JPhCS.840a2021G, 2017PhRvD..95j3012B}. Furthermore, LISA would constrain the secondary mass $M_{\rm sec}$ to within fractional errors of $\lesssim 10^{-3}$, strongly distinguishing between the various possible secondary objects.

Joint observations would also subject the orbital geometry and the central potential to stringent consistency tests. QPE timing models rely on relativistic precession rates (both apsidal and Lense-Thirring) to explain modulations in flare arrival times. Yet, these estimates often suffer from strong degeneracies between the MBH spin $a_{\bullet}$ and orbital eccentricity $e$ \citep{2025MNRAS.543.1816Z}. A coincident GW detection would pin down the MBH spin with an absolute uncertainty of $\Delta a_{\bullet} \sim 10^{-3}$ and the orbital eccentricity to $\Delta e \sim 10^{-4}$ \citep{2017JPhCS.840a2021G, 2017PhRvD..95j3012B}. This precision allows for a direct validation of the specific QPE generation mechanism; for instance, the ``long-short'' recurrence pattern observed in sources like GSN 069 implies a specific eccentricity (e.g., $e \sim 0.1$--$0.2$) \citep{2025arXiv250807961L}, a value that would be confirmed by the relative power of higher-order harmonics in the GW waveform.

In this context, it is also interesting to note that observing the EMRI through electromagnetic messengers would alleviate the LISA data analysis problem. An electromagnetic counterpart would allow for a much more precise sky location than the best estimate obtained from LISA. Additionally, the QPE properties could also provide at least broad Bayesian priors on the other parameters of the system. As such, this total set of priors could be used to narrow the search for the GW signal and, therefore, improve the precision of the other recovered parameters \citep{2014ApJ...795...43F,2017ApJ...834..154P}.

%%%%%%%%%%%%%%%%%%%%%%%%%%%%%%%%%%%%%%%%%%%%%%%%%%%%%%%%%%%%%%%%%%%%%
\subsection{Accretion disk properties} \label{sec:accretiondisk}
%%%%%%%%%%%%%%%%%%%%%%%%%%%%%%%%%%%%%%%%%%%%%%%%%%%%%%%%%%%%%%%%%%%%%

An additional important aspect of joint observations is that they could provide insights into the environment of EMRIs. An EMRI that produces QPEs does not evolve in vacuum, and its GW signal will generally carry imprints of its interaction with the surrounding accretion disk. Beyond the energy losses from GW emission, the disk can introduce additional sources of orbital energy and angular momentum loss. These arise primarily from hydrodynamical drag (see Section \ref{sec:testEMRIQPE}) and dynamical friction when the secondary excites spiral density waves.

These environmental interactions cause the orbit to decay faster than predicted in vacuum, leading to an additional ``dephasing'' of the GW signal relative to a standard template. 
The total phase evolution, $\Phi(t)$, can be expressed as:
\begin{equation}
    \Phi(t) = \Phi_{\rm GW}(t) + \delta\Phi_{\rm disk}(t)
\end{equation}
where $\Phi_{\rm GW}(t)$ represents the evolution due to GW emission alone, and $\delta\Phi_{\rm disk}(t)$ is the accumulated phase shift from disk interactions. This dephasing serves as a direct probe of the disk's properties at the orbiter's location. 
Specifically, measuring this phase shift could allow for the determination of the disk's surface density and viscosity \citep{2023PhRvX..13b1035S,2025ApJ...985..242Z}. 

That being said, as detailed in Section \ref{sec:testEMRIQPE}, these environmental effects are expected to be negligible for compact objects ``punching'' through the disk, unless the orbit is embedded or at a very low inclination \citep{narayan2000hydrodynamic,barausse2014can}. Consequently, looking for and constraining these effects may further clarify the QPE mechanism, distinguishing between high-inclination impacts and the grazing scenarios sometimes invoked to explain high QPE luminosities. Furthermore, the embedded orbiters significantly affected by dynamical friction could instead appear as QPOs (see Section \ref{sec:QPOunification}). If such effects were found, the orbit and disk properties inferred from GW dephasing would have to align with those required to produce the observed electromagnetic QPEs, thus providing a powerful consistency check \citep{2025arXiv250807961L}.

%%%%%%%%%%%%%%%%%%%%%%%%%%%%%%%%%%%%%%%%%%%%%%%%%%%%%%%%%%%%%%%%%%%%%
\subsection{Unveiling EMRI formation channels} \label{sec:EMRIformation}
%%%%%%%%%%%%%%%%%%%%%%%%%%%%%%%%%%%%%%%%%%%%%%%%%%%%%%%%%%%%%%%%%%%%%

As previously mentioned, the eccentricity distribution of EMRIs is a key tracer of their formation history. QPEs and other nuclear transients such as QPOs may already be providing a first, likely heavily biased glimpse into this distribution. The current sample suggests a preference for low-to-moderate eccentricities, favoring ``wet'' formation channels within AGN disks or binary disruption mechanisms \citep{2024PhRvD.110h3019Z}.

LISA is expected to detect hundreds of EMRIs, the vast majority of which are not expected to have EM counterparts. This will provide a ``background'' population distribution of EMRI parameters. By comparing the properties of the EMRI sub-population with EM counterparts (be it a QPE or another nuclear transient) with the full LISA population, we can answer fundamental questions such as:
\begin{itemize}
    \item To which degree are QPEs and other nuclear transients such as QPOs a biased tracer of EMRIs? Do they only occur for EMRIs that happen to interact with a disk, which might select for specific orbital configurations (e.g., low inclination)?
    \item What fraction of EMRIs are ``wet'' (that is, formed by migration in an AGN disk some time in the past \citep{Pan2021}) versus ``dry'' (that is, produced by two body relaxation \citep{Hopman2005} or the Hills' mechanism \citep{Miller2005})?
    \item How does the EMRI formation rate depend on host galaxy properties?
\end{itemize}
It is possible that no EMRI detection will occur by both LISA and in the EM bands for reasons that were outlined in the previous sections of this chapter. In that case, one can still explore the hypothesis that QPEs, other nuclear transients, and LISA EMRIs sample different subpopulations of an ensemble of inspirals formed by a shared set of mechanisms. 

The fact that QPE hosts are often post-starburst, post-merger galaxies with signs of faded AGN activity provides a rich context. This suggests a scenario where galaxy mergers trigger AGN activity, which in turn boosts the formation rate of low-eccentricity EMRIs. The AGN then fades, and a subsequent TDE provides the transient disk needed to ``light up'' the pre-existing EMRI as a QPE \citep{2025ApJ...983L..18J, 2025ApJ...978L...7C}. 

In this picture, the QPE formation rate in a given galaxy is set by the joint probability of forming a TDE disk during the inspiral of an EMRI, such that their crossings will produce observable flares. One can write \citep{2026arXiv260302302A}
\begin{equation}
    R_{\rm QPE} = R_{\rm TDE} \, R_{\rm EMRI} \, t_{\rm GW} \, \mu_{\rm obs} 
\end{equation}
where the TDE rate $R_{\rm TDE}$ and the EMRI rate $R_{\rm EMRI}$ are multiplied by the typical coalescence time of EMRIs $t_{\rm GW}$ set by Equation \eqref{eq:tGW}, and the fraction of EMRIs with orbital parameters producing observable flares $\mu_{\rm obs}$. Because the TDE disk is short-lived compared to $t_{\rm GW}$, the TDE rate acts as a bottleneck, naturally explaining why QPEs preferentially trace environments with enhanced TDE rate. This coupling implies that QPE hosts might be triggered by recent dynamical activity that boosts loss-cone refilling, but the restriction to systems producing observable flares might bias the sample of their host galaxies. Joint EM-GW observations will be crucial for testing this exciting unified narrative, which currently cannot be directly proven.

%%%%%%%%%%%%%%%%%%%%%%%%%%%%%%%%%%%%%%%%%%%%%%%%%%%%%%%%%%%%%%%%%%%%%
\subsection{Cosmology with bright sirens} \label{sec:sirens}
%%%%%%%%%%%%%%%%%%%%%%%%%%%%%%%%%%%%%%%%%%%%%%%%%%%%%%%%%%%%%%%%%%%%%

A GW detection of an inspiraling binary provides a direct measurement of its luminosity distance, $D_{\rm L}$ because the GW strain amplitude is inversely proportional to $D_{\rm L}$, and the absolute (source-frame) amplitude can be determined from the frequency evolution. However, the cosmological redshift $z$ of the source is impossible to recover from the GW analysis alone, since only combinations of redshifts and masses, such as $(1+z) M_\bullet$, appear in the waveform. If an EM counterpart can be identified, the host galaxy's redshift can be measured from its optical spectrum. The combination of $D_{\rm L}$ and $z$ for a single source provides a measurement of the Hubble Constant, $H_0$ \citep{2005ApJ...629...15H}, a feat achieved by the detections of the celebrated binary neutron star merger event GW170817 \citep{2017Natur.551...85A}. These sources are known as ``bright sirens''.

QPE-EMRIs represent ideal bright sirens. Because the QPE phenomenology provides an unambiguous identification of the host galaxy, these systems allow for a direct measurement of the Hubble constant, $H_0$, that is entirely independent of the cosmic distance ladder and its associated systematics. This independence is particularly crucial in light of the ``Hubble tension'', the unexplained $\sim 10\%$ discrepancy between early-universe and late-universe measurements \citep{2016A&A...594A..13P,2018ApJ...855..136R,riess2024jwst}. While a single QPE-EMRI detection by LISA could constrain $H_0$ with a precision of $\sim 10 \%$ \citep{Zhan2026}, combining observations from just a handful of such sources could refine this to the few-percent level, providing a useful new constraint to help resolve the tension.

%%%%%%%%%%%%%%%%%%%%%%%%%%%%%%%%%%%%%%%%%%%%%%%%%%%%%%%%%%%%%%%%%%%%%
%%%%%%%%%%%%%%%%%%%%%%%%%%%%%%%%%%%%%%%%%%%%%%%%%%%%%%%%%%%%%%%%%%%%%
\section{The future: The need for a coordinated search strategy} \label{sec:future}
%%%%%%%%%%%%%%%%%%%%%%%%%%%%%%%%%%%%%%%%%%%%%%%%%%%%%%%%%%%%%%%%%%%%%
%%%%%%%%%%%%%%%%%%%%%%%%%%%%%%%%%%%%%%%%%%%%%%%%%%%%%%%%%%%%%%%%%%%%%

The prospect of multimessenger astronomy with QPE-EMRIs calls for a coordinated strategy between current and future EM observatories and the LISA mission since, in order to be detected by LISA, QPEs need periods of minutes instead of hours as is the case for the currently known QPEs. We start by laying out the question of how could one detect a short-period ``golden'' QPE in Section \ref{sec:golden}. This leads to the conclusion that we first need to expand the knowledge of QPE populations mainly by expanding the catalog of QPEs and searching for evidence to support and constrain the EMRI model, which we discuss in Section \ref{sec:expandcatalog}. Finally, in Section \ref{sec:operationsynergy} we discuss the synergies that will arise as a result of a coordinated effort when LISA becomes operational.

%%%%%%%%%%%%%%%%%%%%%%%%%%%%%%%%%%%%%%%%%%%%%%%%%%%%%%%%%%%%%%%%%%%%%
\subsection{How to find a ``golden'' QPE?} \label{sec:golden}
%%%%%%%%%%%%%%%%%%%%%%%%%%%%%%%%%%%%%%%%%%%%%%%%%%%%%%%%%%%%%%%%%%%%%

\begin{figure}
    \centering
    \includegraphics[height=0.38\linewidth]{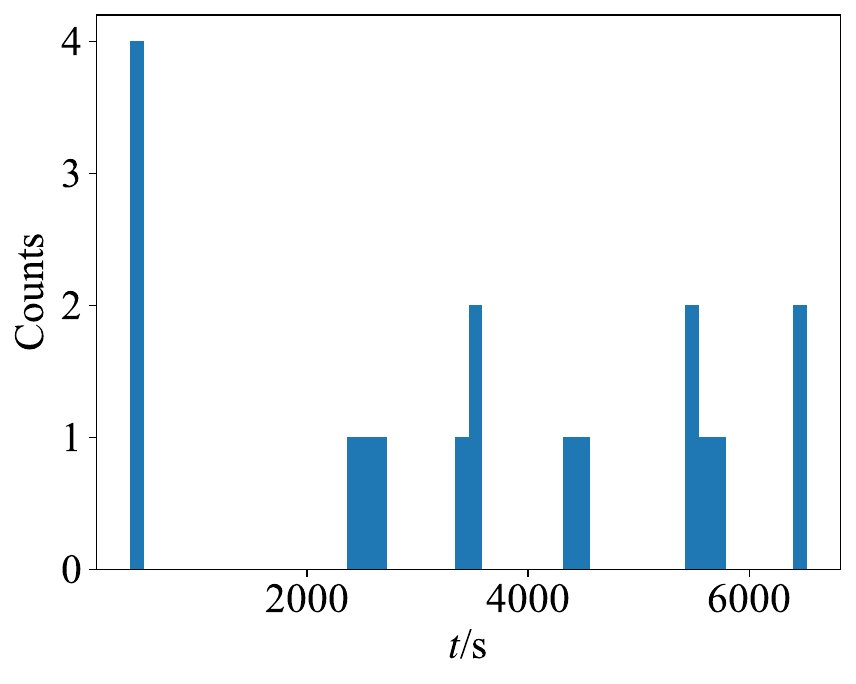}
    \includegraphics[height=0.38\linewidth]{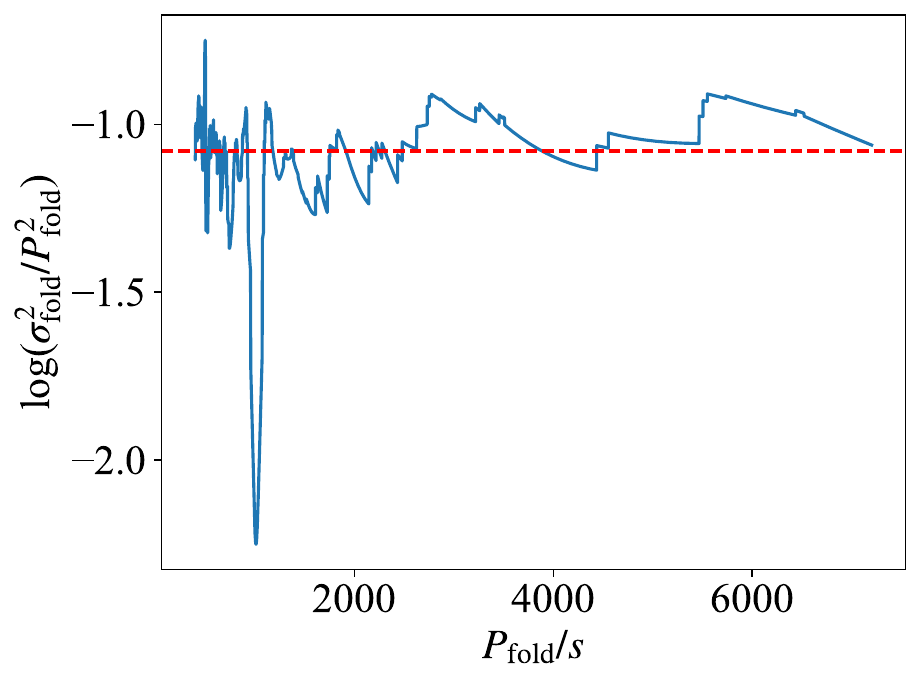}
    \caption{\textit{Left:} A toy model for X-ray photon detections of a QPE with a short, 1000 second period. Single eruptions correspond to units of photons detected, with a total of 18 photons detected in 7200 seconds of observation. The second eruption at $t=1500$ s is also present but the count rate is so low that no photon arrived in the Poisson-process simulation. \textit{Right:} The variance $\sigma_{\rm{fold}}^2$ of the folded epochs of the photons from the left Figure normalized by folding period $P_{\rm fold}$ squared plotted as a function of folding period. The dashed red line corresponds to the mean variance of a homogeneous Poisson process for reference. The true period of the QPE is robustly identified as the minimum of the blue curve.}
    \label{fig:ShortQPE1}
\end{figure}

\citet{Kejriwal2024} argued that repeating nuclear transients with periods in the range of tens of minutes are the most likely to drift towards frequencies in the LISA band and generate measurable EMRI signals by the late 2030s. How can we search for or even detect such ``golden'' QPEs? 

Regarding the purely technological detection feasibility, the challenge and its potential resolution through epoch-folding methods \citep{stellingwerf1978period,schwarzenberg1989advantage,ivezic2020statistics} can be illustrated with a simple toy model as follows. The fundamental issue is that even if we assume that the peak luminosities of the short-period QPEs are the same as those of the known long-period QPEs, we must assume a similar $\sim 10-30\%$ duty cycle. Assuming a $\sim$ 1000 second QPE, this leads to $\sim 100$ second eruptions. Peak photon count rates for known QPEs are $\sim$ few per second. However, known QPEs are mostly within $\sim 200$ Mpc while LISA will detect EMRIs within a few Gpc \citep[see, e.g., Fig. 3 of][]{Kejriwal2024}. It is thus interesting to ask whether short-period QPEs would also be detectable to such ``typical'' LISA EMRI distances. Assuming the same peak luminosity as known QPEs but at a few Gpc, this would lead to peak count rates of $\sim 10^{-2}$ per second with current instruments. Along with the assumed eruptions durations, this implies that the integrated number of photons detected per eruption may be of the order of units. Conversely, even a few kiloseconds of observations should capture a number of periods, allowing for a periodicity search through epoch folding techniques. 
    
As an illustration, we generated a simulated detection of a 1000 second QPE with a 10\% duty cycle, peak flux of $10^{-2}$ photons per second, and negligible background flux. The observation is assumed to run for 7200 seconds, and the random arrival times of the photons are modeled as an inhomogeneous Poisson process with the arrival probability of a photon modulated by the flux intensity. 
We obtained a total of 18 photons, and we show the binned photon counts as a function of time in the left panel of Fig. \ref{fig:ShortQPE1}. In order to understand how many cycles are necessary to correctly identify the periodicity, we fold the detection epochs (photon arrival times) with a test period $P_{\rm fold}$ and search for the minimum of the variance of the folded arrival times as a function of $P_{\rm fold}$. The idea of this technique is that when a test period equal to the underlying QPE period of 1000 seconds is used, the arrival times cluster around a single peak corresponding to the folded eruptions, and the dispersion of the folded arrival times is low. However, for a test folding period unrelated to the recurrence time of the eruptions, arrival times from different eruptions will not add up coherently, and the dispersion $\sigma_{\rm fold}$ will generally be close to the value of a homogeneous Poisson process, which is $P_{\rm fold}/\sqrt{12}$. We show the variance $\sigma^2_{\rm fold}$, normalized by $P^2_{\rm fold}$, as a function of $P_{\rm fold}$ in the right panel of Fig. \ref{fig:ShortQPE1}. We see that the period of the QPE (i.e., 1000 seconds) can be robustly recovered even from just the 18 detected photons in this toy model \citep[for more on these methods, see][]{stellingwerf1978period,schwarzenberg1989advantage,ivezic2020statistics}. 

Furthermore, an evaluation of $10^5$ Monte Carlo runs indicates that the false alarm probability for detecting this period in a homogeneous Poisson process producing the same number of photons at random times is less than $\sim 6.5 \cdot 10^{-7}$ (or $\sim 5 \sigma$ from the mean). Generally, such illustrative results suggest that observational windows shorter than $\sim 7$ periods of the QPE do not detect the period robustly. This implies that one should be able to use $\sim 10$-ks observation windows or longer per candidate source when searching for LISA-relevant counterparts. This is in line with the durations of typical single observations by the \textit{Chandra} or \textit{XMM-Newton} X-ray observatories.     
    
While the promise of detecting such a short-period QPE is immense from the perspective of multi-messenger astronomy, it is also a very risky proposal from the perspective of astrophysical likelihood. This is because QPEs are currently mostly understood only phenomenologically and do not have well-established formation and population models. Assumptions about the duty cycle or peak luminosities of the QPEs, while valid for known low-frequency QPEs, may also become invalid in the mHz range, further complicating the prospects of such a campaign. 

For example, a QPE EMRI in this period range would correspond to either a stellar object very close to tidal disruption or a compact object causing the eruptions. Given the short lifetime of the stellar object on such an orbit, the likelihood of finding a corresponding QPE would be very small, even with a large-scale survey. In the case of a compact object, it would likely need to be an IMBH or a grazing stellar-mass BH to affect enough matter in the disk and generate a large enough eruption (see Section \ref{sec:luminosity}). As seen in eq. \eqref{eq:tGW}, the GW inspiral time of such binaries scales with the orbital period as $P^{8/3}$, so it is intrinsically less likely to find a short-period binary in a galactic nucleus than a long-period one \citep[see also][]{2026arXiv260302302A}. Assuming that the population of golden QPEs is the same as the one observed currently, just evolved forward in time, we can estimate that we would observe the golden QPEs with a likelihood a few orders of magnitude lower than that of the currently observed QPEs. In conclusion, while a golden QPE would provide an interesting target for LISA, one should generally expect such phenomena to be very rare.    

As a result, even an observation campaign involving tens or hundreds of candidate sources, which would require significant resources in terms of observation time, would have an unquantifiable and possibly very low chance of success. This means that even though a search for such sources would be ``high reward'', it would also be too ``high risk'' with the current gaps in our knowledge. We thus need to first expand the known populations of QPEs, narrow down the theoretical scenarios, and build a formation model. This will enable us to identify a more promising set of candidate sources before possibly launching a well balanced high-risk/high-reward observation campaign for short-period QPEs.

%%%%%%%%%%%%%%%%%%%%%%%%%%%%%%%%%%%%%%%%%%%%%%%%%%%%%%%%%%%%%%%%%%%%%
\subsection{Expanding QPE catalogs to bridge the gap} \label{sec:expandcatalog}
%%%%%%%%%%%%%%%%%%%%%%%%%%%%%%%%%%%%%%%%%%%%%%%%%%%%%%%%%%%%%%%%%%%%%

In order to mitigate the risk of blind searches for short-period sources, we must first constrain the population and emission models by expanding our knowledge of the more accessible long-period population. 

The first step is to expand the sample of QPEs. All-sky X-ray surveys like eROSITA have been instrumental in finding new QPEs \citep{2024A&A...684A..64A}. Future high-cadence monitoring missions, such as the proposed Chinese CATCH mission (a proposed constellation of microsatellites for X-ray timing), will be crucial for discovering dozens more and for obtaining the long-baseline timing data needed to measure orbital evolution \citep{2025MNRAS.543.1816Z}.
    
There are also additional classes of QPEs that one can search for. Theoretical models predict that as a TDE disk evolves and cools, the QPEs produced by star-disk collisions may shift from the X-ray to the ultraviolet (UV) \citep{2024ApJ...963L...1L}. This transition occurs because, at lower accretion rates, the shock-heated debris achieves thermal equilibrium (dropping the flare temperature to the UV range), while the quiescent disk simultaneously dims sufficiently to reveal these softer eruptions. Future UV missions like ULTRASAT \citep{2024ApJ...964...74S} could open a new window on this phenomenon, potentially discovering a population of ``UV QPEs'' and tracking the late-stage evolution of these systems.

While the current population of QPEs is primarily detected years after the peak emission of a TDE when the disk has transitioned into a sub-Eddington, geometrically thin state, the secondary objects could also interact with a ``slim disk'' formed by the early-TDE evolution.  Specifically, collisions with the high-density, radiation-dominated environment of the early TDE disk are predicted to yield eruptions that are significantly harder ($1\text{--}100$~keV) and shorter ($\sim 10^2\text{--}10^3$~s) than those currently observed while exhibiting similar recurrence times \citep{2025MNRAS.tmp.2110S}. Confirming or disproving such models would provide much needed differentiation between the various QPE scenarios and thus also much tighter constraints on the possible nature of the secondary, clarifying the possibility of a coincident GW detection.

%%%%%%%%%%%%%%%%%%%%%%%%%%%%%%%%%%%%%%%%%%%%%%%%%%%%%%%%%%%%%%%%%%%%%
\subsection{Operational synergy in the LISA era and beyond} \label{sec:operationsynergy}
%%%%%%%%%%%%%%%%%%%%%%%%%%%%%%%%%%%%%%%%%%%%%%%%%%%%%%%%%%%%%%%%%%%%%

Once LISA is operational, the catalog of known QPEs will serve as a priority list for targeted GW searches. This could provide EMRI detections extracted from the data stream earlier than those found by blind searches. Still, only a fraction of QPEs might result in detectable LISA EMRI sources, given the weakness of the signal as the QPE period increases. Hierarchical population analysis will be crucial to navigate the multiple biases that the QPE-EMRI connection brings.

Conversely, as the LISA data stream is analyzed, candidate EMRI events will be identified. These will come with sky localization boxes. EM observatories can then be triggered to search these regions for transient phenomena, potentially discovering QPEs (or other counterparts) that were previously missed. This two-way synergy will maximize the scientific return.

However, as discussed in the previous sections, the current sample of QPEs, when interpreted as EMRIs, might have only a couple of sufficiently loud and thus detectable sources. If longer period QPEs represent the bulk of the population, such EMRI abundance might emit GWs at too low a frequency for LISA. Possible future GW observatories targeting a lower frequency window  could reveal more about the underlying physical landscape previously concealed beneath the most evident phenomena. Proposed detector concepts for such observatories involve GW detectors spanning the Earth or Mars orbit, or detectors placed on asteroids \citep{2021ExA....51.1333S,fedderke2022asteroids,2023CQGra..40s5022M}.

For example, the more conservatively formulated LISAmax concept proposes a triangular constellation with arm lengths of 260 million kilometers located at the Sun-Earth Lagrange points L3, L4, and L5 \citep{2023CQGra..40s5022M}. This configuration shifts the observatory's primary sensitivity window to frequencies approximately an order of magnitude lower than LISA, effectively opening the $\mu$Hz frequency band for observation. For a QPE driven by a binary with an $\sim 10$-hour orbital period, the resulting gravitational-wave signal would occur at $60 \mu$Hz. While this frequency lies in a regime where LISA's sensitivity degrades rapidly, it falls squarely within the high-sensitivity region of the LISAmax design. Consequently, LISAmax would be uniquely positioned to detect these longer-period inspirals, potentially revealing the population of EMRIs responsible for QPEs well before they evolve into the frequency band accessible to LISA.

%%%%%%%%%%%%%%%%%%%%%%%%%%%%%%%%%%%%%%%%%%%%%%%%%%%%%%%%%%%%%%%%%%%%%
%%%%%%%%%%%%%%%%%%%%%%%%%%%%%%%%%%%%%%%%%%%%%%%%%%%%%%%%%%%%%%%%%%%%%
\section{Conclusion}
%%%%%%%%%%%%%%%%%%%%%%%%%%%%%%%%%%%%%%%%%%%%%%%%%%%%%%%%%%%%%%%%%%%%%
%%%%%%%%%%%%%%%%%%%%%%%%%%%%%%%%%%%%%%%%%%%%%%%%%%%%%%%%%%%%%%%%%%%%%

Quasi-Periodic Eruptions have evolved from a celestial curiosity into a potential cornerstone of multimessenger astrophysics. The evidence possibly linking them to Extreme Mass-Ratio Inspirals places them at the intersection of accretion physics, dynamics in galactic nuclei, and GW science. If the EMRI model holds, QPEs could become the electromagnetic herald of a new era of GW astronomy, offering a direct path to the first confident detections of EMRIs with LISA. 

Nevertheless, all known QPEs exhibit recurrence periods that are too long for the corresponding GW signal to fall within the optimal sensitivity band of the Laser Interferometer Space Antenna (LISA), which will be most sensitive to GWs sourced by orbital phenomena with $\sim$10-minute to 10-second periods. Furthermore, a notable population disparity exists: while QPE luminosities seem to favor main-sequence stellar secondaries, LISA will be biased toward heavier compact objects, such as black holes, which also do not undergo tidal disruption before reaching millihertz frequencies.

If these observational gaps are bridged through the discovery of shorter-period QPEs, the scientific payoff would include precise measurements of black hole mass and spin that would be linked to their host galaxies. Such multimessenger systems could serve as ``bright sirens" for cosmology, providing independent estimates of the Hubble constant to help resolve current measurement tensions. Achieving this requires developing a multi-stage coordinated strategy to identify transients with periods in the range of tens of minutes and developing robust population models before LISA's mission begins in the late 2030s.

The path forward requires a dedicated, synergistic effort. Continued X-ray monitoring is needed to expand the QPE catalog and to track their long-term evolution, testing the predictions of the EMRI model. These EM observations will provide the essential ``treasure maps'' for targeted searches in the LISA data and population models that will allow us to search for short-period QPEs. The subsequent joint analysis of EM and GW data would unlock a new level of understanding of the dynamics of galactic nuclei, the growth of supermassive black holes, and the fundamental nature of gravity in the strong-field regime. 

\bibliography{papers}

\end{document}